\documentclass[a4paper,11pt]{article}
\pdfoutput=1 

\usepackage{jcappub} 

\usepackage[T1]{fontenc} 
\usepackage{graphicx}
\newcommand{\hmpc}{{\, h^{-1}\, {\rm Mpc}}}

\def\aj{AJ}
\def\apj{ApJ}

\def\mnras{MNRAS}

\def\nat{Nature}

\title {\boldmath Green valley galaxies in the cosmic web: internal versus environmental quenching}

\author[a]{Apashanka Das,}

\author[a]{Biswajit Pandey,}

\author[b]{and Suman Sarkar}
\affiliation[a]{Department of Physics, Visva-Bharati University,
   Santiniketan, 731235, India}
\affiliation[b]{Department of Physics, Indian Institute of Science Education and Research Tirupati, Tirupati - 517507, Andhra Pradesh, India}
\emailAdd{a.das.cosmo@gmail.com} \emailAdd{biswap@visva-bharati.ac.in} \emailAdd{suman2reach@gmail.com}

  \abstract{We analyze the SDSS data to classify the galaxies based on their
  colour using a fuzzy set-theoretic method and quantify their
  environments using the local dimension. We find that the fraction of
  the green galaxies does not depend on the environment and
  $10\%-20\%$ of the galaxies at each environment are in the green
  valley depending on the stellar mass range chosen. Approximately
  $10\%$ of the green galaxies at each environment host an
  AGN. Combining data from the Galaxy Zoo, we find that $\sim 95\%$ of
  the green galaxies are spirals and $\sim 5\%$ are ellipticals at
  each environment. Only $\sim 8\%$ of green galaxies exhibit signs of
  interactions and mergers, $\sim 1\%$ have dominant bulge, and $\sim
  6\%$ host a bar. We show that the stellar mass distributions for the
  red and green galaxies are quite similar at each environment.  Our
  analysis suggests that the majority of the green galaxies must
  curtail their star formation using physical mechanism(s) other than
  interactions, mergers, and those driven by bulge, bar and AGN
  activity. We speculate that these are the massive galaxies that have
  grown only via smooth accretion and suppressed the star formation
  primarily through mass driven quenching. Using a Kolmogorov-Smirnov
  test, we do not find any statistically significant difference
  between the properties of green galaxies in different
  environments. We conclude that the environmental factors play a
  minor role and the internal processes play the dominant role in
  quenching star formation in the green valley galaxies.}

\begin{document}
\maketitle
\flushbottom


\section{Introduction}

One of the most coveted goals of cosmology is understanding the
formation and evolution of galaxies in the Universe. Galaxy surveys
(e.g. 2DFGRS \citep{colless01}; SDSS \citep{strauss02}) over the last
few decades have measured the properties of millions of galaxies to
unprecedented accuracy. The observed galaxy properties such as colour
\citep{strateva01, hogg03, balogh04, baldry04}, star formation rate
and stellar age \citep{kauffmann03a}, bulge to disc ratio
\citep{kauffmann03a}, gas to stellar mass ratio \citep{kannappan04}
show a bimodal character. The galaxy colour is relatively easy to
calculate. It characterizes the stellar population of a galaxy. The
distribution of the optical colour show two distinct peaks
corresponding to two galaxy populations which are often referred to as
the `red sequence' and the `blue cloud'. The two population show
significant overlap and the relatively shallow and flat region between
the two peaks is termed as `green valley' \citep{wyder07}. The
galaxies in the blue cloud are predominantly the star-forming
population with disk-like morphology, younger stellar population and
lower stellar mass whereas the galaxies in the red sequence are
primarily the quiescent galaxies with a dominant bulge component,
older stellar population and higher stellar mass \citep{strateva01,
  kauffmann03a, blanton03, baldry04}. The morphology of a galaxy
correlates well with its stellar population. However, morphological
transformations alone, can not explain the bimodal distribution of
galaxy colour. Studies with Galaxy Zoo \citep{lintott08} reveal that a
significant number of ellipticals are contained in the blue cloud
\citep{schawinski09a} and a large number of spirals are found in the
red sequence \citep{masters10}.

Understanding the origin of the bimodal nature of the colour
distribution and its evolution holds the important keys to galaxy
formation and evolution. Different semi-analytic models have been used
to reproduce the observed bimodal distribution of galaxy colour
\citep{menci,driver,cattaneo1,cattaneo2,cameron,trayford16,nelson,correa19}
and gain insight into the physical processes leading to
bimodality. Observations suggest that the bimodality in galaxy colour
exists out to a redshift of $z=1-2$ \citep{bell04b, weiner05, kriek08,
  brammer09}. Besides, the number of massive red galaxies at fixed
stellar mass has been found to increase steadily since $z \sim 1$
\citep{bell04b, faber07}. The steady increase in the number of red
galaxies and their stellar mass indicates that this could happen if
the galaxies from the blue cloud migrate to the red sequence via
quenching of star formation. The increase in the red stellar mass can
be also achieved by the merger of already quenched less massive
galaxies. Further, it has been reported that the cosmic star formation
shows a sharp decline between $z=1$ to $z=0$ \citep{madau96}. These
observations point towards a significant evolution in the galaxy
properties in the recent past, which may have played a pivotal role in
the observed bimodality of their distributions.

The current paradigm of galaxy formation owes its root to a few
seminal papers from the seventies \citep{reesostriker77, silk77,
  white78, fall80}. The baryons are believed to be shock heated to the
halo virial temperature after they fall into the dark matter potential
well. The baryons in the inner region of the halo then radiatively
cool and settle down on a dynamical time scale to form rotationally
supported disk galaxies. These galaxies inside the dark matter halos
then evolve either in isolation or through the interactions with their
environment. Both the nature (secular) versus nurture (environment)
scenario of galaxy evolution, are capable of producing observed
bimodality in the galaxy colour distribution.

The secular evolution of a galaxy is not limited to mere aging of its
stellar population and there are many internal physical processes
(e.g. mass quenching, morphological quenching, bar quenching) which
can play a significant role in the reduction of star formation in a
disk galaxy. Kauffmann et al. \citep{kauffmann03a} find that the
galaxies with a mass less than $3 \times 10^{10}\,M_{\odot}$ show
active star formation, lower surface mass density and disk-like
morphology whereas those with masses larger than this critical value
are generally quiescent galaxies with bulge dominated morphology and
higher surface mass density. This critical stellar mass corresponds to
a critical halo mass of $10^{12}\,M_{\odot}$ which can be associated
with the observed bimodality. The theoretical analysis
\citep{binney04} and studies with hydrodynamical simulations
\citep{birnboim03, dekel06, keres05, gabor10,gabor15} suggest that
this critical mass is associated with a transition from `cold mode' to
`hot mode' of accretion. Simulations show that the cooling time is
shorter than the dynamical time in halos with mass less than
$10^{12}\,M_{\odot}$ and they can accrete cold gas through
quasi-spherical filamentary inflows maintaining their star
formation. But in the massive halos, cooling time is way longer where
a stable shock expands to the virial radius producing a hot medium at
the virial temperature, which prevents any cold streams from the
inter-galactic medium to penetrate through the hot gas without getting
heated. So the virial shock heating of the halo gas in massive halos
with mass greater than $10^{12}\,M_{\odot}$ can suppress the supply of
cold gas required for star formation. This quenching of star formation
in high mass halos is known as the `mass quenching'. We refer to this
halo quenching as mass driven quenching throughout this
work. Simulations suggest that the mass quenching alone can not
maintain long-term shutdown of star formation as the shock heated hot
gas eventually cool down via radiation and then collapse at the
centres of these halos to form stars \citep{birnboim07}. An additional
heating source such as radio mode AGN feedback can prevent such
cooling and maintain the high temperature of the halo gas
\citep{croton06, bower06, somerville08}.  The gravitational heating
due to clumpy accretion at the centre of the halos
\citep{birnboim07,dekel08,dekel09b} can also prevent cooling and star
formation. So a coupling of the virial shock heating with these
additional energy feedback processes can shutdown the star formation
in dark matter halos above the critical mass.

The `bar quenching' and `morphological quenching' are some of the
other internal processes which can shut down the star formation in a
galaxy. The presence of stellar bars in disk galaxies can suppress star
formation \citep{haywood16,spinoso17,james18,george19}. The
bar-induced torque can transfer gas from the outskirts to the centre
of the galaxy leading to the buildup of a bulge
\citep{combes81,debattista04,kormendy04,athanassoula13}. The central
bulge produces a deeper potential well which can stabilize the disk
against collapse \citep{martig09}. The presence of the bulge is also
known to trigger AGN activity \citep{bruce16} and nuclear star
formation turning the inner kilopersec region devoid of any cold gas
\citep{combesgerin85,fang13,spinoso17}.

Besides these internal processes, the environment of a galaxy can also
play a significant role in the quenching of star formation. Major
mergers of galaxies can transform spirals to ellipticals
\citep{toomre72, barnes02}. They can also facilitate the ejection of
interstellar medium through starburst and AGN or shock-driven winds
\citep{cox04,murray05,springel05} thereby quenching star formation in
those galaxies. Some of the other important routes for environmental
quenching are the ram pressure stripping of cold gas \citep{gunn72},
galaxy harassment in clusters \citep{moore96,moore98}, strangulation
in galaxy-group interactions \citep{gunn72,balogh00} and starvation
due to truncation of gas supply \citep{larson80,somerville99,kawata08}.

The blue galaxies evolve to red galaxies via quenching of star
formation.  The galaxies with intermediate properties between blue and
red galaxies lie in the green valley undergoing such a transition. It
is important to understand the primary physical mechanisms responsible
for such transition. The red galaxies are preferably found in the
denser region \citep{hoyle02, park05} which suggests the role of
environment in quenching.  Some studies suggest that AGN feedback
plays an important role in quenching star formation in the
transitional green valley population
\citep{nandra07,hasinger08,silverman08,cimatti13}.  The analysis of
Galaxy Zoo by Schawinski et al.\citep{schawinski14} suggests that
quenching in green valley galaxies require the gas reservoir
destruction caused by starburst and AGN feedback in major
mergers. Alternatively, starving galaxies of the cold gas supply and
exhaustion of the remaining gas can also initiate such quenching. Lin
et al. \citep{lin17} find lower star formation efficiency in the green
valley galaxies from the study of their cold molecular gas content,
which indicates that complete removal of cold gas supply is not
necessary for quenching in these galaxies. A comprehensive review of
the green valley galaxies can be found in Salim \citep{salim}. Coenda
et al. \citep{coenda18} study the properties of green valley galaxies
in fields, groups and clusters using SDSS and find that there is a
clear environmental dependence of external quenching mechanisms in the
green valley galaxies. Jian et al. \citep{jian20} study the redshift
evolution of green valley galaxies in different environments to find a
mild evolution in the environmental dependence which suggests that
slow quenching mechanisms are operating in denser environments since
$z \sim 1$.

The timescales associated with the different quenching mechanisms vary
widely spanning from few hundred Myr to 1 Gyr or more. The
morphological quenching, bar quenching and strangulation belong to
slow modes of quenching which can extend to 1-2 Gyr
\citep{martig09,larson80,balogh00} whereas the mergers and ram
pressure stripping can halt the star formation on a time scale of less
than 1 Gyr \citep{gunn72, mihos94}. An analysis of the EAGLE
cosmological simulation by Trayford et al. \citep{trayford16} suggests
that the galaxies require $\sim 2$ Gyr to cross the green valley
regardless of the quenching mechanism.

The well known morphology-density relation
\citep{hubble36,dressler80,postman84} and SFR-density relation
\citep{lewis02,gomez03,kauffmann04} suggest that environment plays an
important role in deciding the galaxy properties and their
evolution. The galaxies are organized in an interconnected network of
filaments, sheets and clusters encircled by voids, which is often
referred to as the ``cosmic web'' \citep{bond96}.  Studies with
cosmological N-body simulations show that the properties such as mass,
shape and angular momentum of dark matter halos are influenced by
their geometric environments in the cosmic web \citep{hahn07}. The
clustering of the dark matter halos in the cosmic web is also known to
depend on their assembly history \citep{croton07, gao07, musso18,
  vakili19}. Several observational studies
\citep{miyatake16,montero17,kerscher18} claim evidence of assembly
bias. Since the cosmic web can influence both the properties and
clustering of dark matter halos, it can also play a significant role
in the formation and evolution of galaxies. Many earlier studies
\citep{pandey06,pandey08,scudder12,darvish14,filho15,luparello15} with
observational data find a significant correlation between galaxy
properties and large-scale environment. Pandey \& Sarkar
\citep{pandey17} analyze data from Galaxy Zoo and find that a synergic
interaction between the morphology of the galaxies and their
large-scale environments persists at least up to $30 \hmpc$. Using
SDSS, Sarkar \& Pandey \citep{sarkar20} show that the mutual
information between morphology and large-scale environment are
statistically significant at $99.9\%$ confidence level. A recent
analysis by Bhattacharjee et al. \citep{bhattacharjee20} show that
conditioning on stellar mass does not provide a complete explanation
of the mutual information between morphology and the large-scale
environment. Pandey \& Sarkar \citep{pandey20a} analyze the data from
SDSS to find that at a fixed density, the fraction of red and blue
galaxies are sensitive to the geometric environments of the cosmic
web.

The SDSS \citep{strauss02} is the largest redshift survey to date. It
provides the photometric and spectroscopic information of millions of
galaxies which allows us to carry out statistical analysis of the data
and address many important issues related to galaxy formation and
evolution. The galaxies in the green valley hold important clues about
galaxy evolution. It is important to understand the role of internal
and external influences on the suppression of star formation in the
green valley galaxies. In the present work, we plan to study the
fraction of green galaxies and their possible routes of quenching in
different geometric environments of the cosmic web. As mass plays an
important role in quenching, the environmental dependence of the
fractions should be investigated at fixed stellar mass. Analysis in
different mass bins may help us to identify the influence of the
environments on quenching. It is also well known that the presence of
a dominant bulge, a bar or AGN activity can quench star formation in a
galaxy \citep{fisher08,fabian12,lang14,forster14,leslie16, haywood16,
  george19}. We would like to investigate the relative importance of
these quenching agents in different environments of the cosmic web. It
is also important to compare the distributions of the stellar mass,
star formation rate and star formation history of the green valley
galaxies in different geometric environments to understand the role of
environments in deciding the properties of green galaxies. We also
address if the length scales associated with the host environments
affect the galaxies in the green valley.

Recently Pandey \citep{pandey20b} propose a fuzzy-set theoretic method
for classification of galaxies based on their colour. We use this
method to select the green valley galaxies for the present
analysis. We quantify the geometric environments of the galaxies using
their local dimension \citep{sarkar09}. Our primary goal in this work
is to understand the relative importance of internal processes and
external environments for the transitional green valley population.

The paper is organised as follows, we describe the data in Section 2,
the method of analysis in Section 3, discuss the results in Section 4
and present our conclusions in Section 5.

\section{SDSS Data}

The Sloan Digital Sky Survey (SDSS) is a multi-band photometric and
spectroscopic redshift survey that uses a dedicated 2.5 m telescope at
Apache Point Observatory in New Mexico to measure the spectra and
images of millions of galaxies over roughly one third of the sky. Gunn
et al. \citep{gunn98} describe the technical details of the SDSS
photometric camera. The construction, design and performance of the
SDSS telescope is described in Gunn et al. \citep{gunn06}. The
algorithm for selecting the SDSS main sample for spectroscopy is
provided in Strauss et al. \citep{strauss02}. A technical summary of
the SDSS is outlined in York et al. \citep{york00}.

We use the publicly available data from SDSS DR16 \citep{ahumada20}
for the present analysis. We retrieve the required data from
\textit{Casjobs} \footnote{https://skyserver.sdss.org/casjobs/} using
\textit{Structured Query Language}. We have downloaded the
spectroscopic and photometric information of galaxies from
\textit{SpecPhotoAll} table 
\begin{figure*}[htbp!]
\centering
\includegraphics[width=12cm]{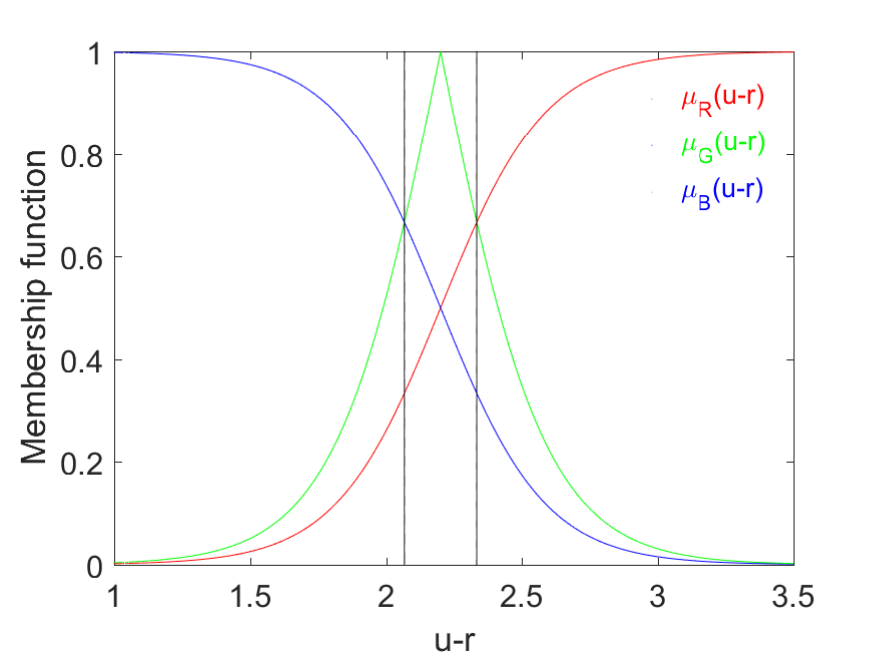}
\caption{This figure shows the definition of the red, blue and green
  galaxies classified using fuzzy set theory. The region between the
  two vertical solid lines corresponds to the green galaxies.}
\label{fig:fuzzycol}
\end{figure*} 
using the following cuts: $z < 0.2$, $13.5\le r_p<17.77$,
$135^{\circ}\le\alpha\le 225^{\circ}$, $0^{\circ}\le\delta\le
60^{\circ}$ where $z$ is the redshift, $r_p$ is the $r$-band Petrosian
magnitude, $\alpha$ and $\delta$ are right ascension and declination
respectively. We also download the stellar mass and star formation
rate of these galaxies from $stellarMassFSPSGranWideNoDust$ table
\citep{conroy09}, K correction from $Photoz$ table and $4000\AA$ break
strength D4000 \citep{bruzal83} from $galSpecIndx$ table
\citep{brinchmann04}. We use $galSpecExtra$ table derived from MPA-JHU
spectroscopic catalogue \citep{brinchmann04, kauffmann03b} to identify
the AGNs classified on the basis of BPT emission line diagram. The
SDSS is a fibre based survey and the aperture effects are important in
the measurement of these spectroscopic properties. These biases have
been properly taken into account in these measurements using aperture
correction schemes described in Brinchmann et
al. \citep{brinchmann04}. The galaxies with AGN activity are flagged
as $4$ in the table $galSpecExtra$. For morphological classification
we use $zooSpec$ table derived from Galaxy Zoo \citep{lintott08}. We
identify the elliptical and spiral galaxies from this table as those
which have their $elliptical$ and $spiral$ flag set to 1 (with
debiased vote fraction $> 0.8$) respectively. Further we have
downloaded the information of internal reddening E(B-V) for each
galaxies from $emmissionLinesPort$ table which is based on publicly
available Gas AND Absorption Line Fitting (GANDALF) \citep{gandalf}
and penalised PIXEL Fitting (pPXF) \citep{ppxf}. We combine these
tables and apply the above mentioned cuts which provides us with
321741 galaxies. We construct a volume limited sample from this data
by restricting the K-corrected and extinction corrected $r$-band
absolute magnitude to $-23\le M_r \le -21$ which corresponds to a
redshift cut of $0.041\le z\le 0.120$. The resulting volume limited
sample consists of 93187 galaxies. Throughout this analysis, we use
$\Lambda$CDM cosmological model with
$\Omega_{m0}=0.315,\Omega_{\Lambda 0}=0.685$ and $h=0.674$
\citep{planck18}.

\section{METHOD OF ANALYSIS}
\subsection{Classifying the red, blue and green galaxies using fuzzy set theory}

We classify the galaxies as red, blue and green using a fuzzy set
theory based method recently proposed in \citep{pandey20b}. The method
is described in detail in \citep{pandey20b}. Here we outline the basic
idea underlying this method.

A fuzzy set $A$ is defined by a set of ordered pairs
\citep{zadeh},
\begin{eqnarray}
A=\big{\{}\,(\,x,\,\mu_{A}(x)\,) \,\,|\,\,x \in X\,\big{\}} 
\label{eq:fuzzy1}
\end{eqnarray}
where $X$ is the Universal set and $\mu_{A}(x)$ is
the membership function of the fuzzy set $A$. The membership function
maps the elements of the Universal set $X$ to real numbers in
$[0,1]$. It provides the possibility of being member of the fuzzy set
$A$. In other words, it measures the degree of membership of a
particular element in the fuzzy set.

  In reality, the galaxies can not be strictly classified as red, blue
  or green based on their colours. Such a classification is solely
  motivated by the bimodal distribution of the galaxy colour. In any
  such classification, it is important to differentiate between the
  redder galaxies with a reduced star-formation activity and that are
  dusty. This requires the observed colours to be corrected for dust
  attenuation. We correct the observed $u-r$ colour using the internal
  reddening E(B-V) for each galaxies.

We define the fuzzy set $R$ corresponding to redness of all galaxies
in the volume limited sample as,
\begin{eqnarray}
R=\big{\{}\, (u-r,\, \mu_{R}(u-r))\,\,|\,\, (u-r) \in X \,\big{\}} 
\label{eq:fuzzy2}
\end{eqnarray}
where $X$ is the Universal set of dust corrected $(u-r)$ colour. The
membership to the fuzzy set is described with a sigmoidal function
\citep{pandey20b},
\begin{eqnarray}
\mu_{R}(u-r;a,c)=\frac{1}{1+e^{-a[(u-r)-c]}}
\label{eq:fuzzy3}
\end{eqnarray}
Here $a$ and $c$ are constants. We choose $a=5.2$ and $c=2.2$ as
specified in \citep{pandey20b}. The choice of the sigmoidal membership
function is based on the bimodal nature of the $u-r$ colour
distribution. The two peaks of the bimodal distribution corresponds to
the `blue cloud' and `red sequence'. The parameters $c$ and $a$
respectively denote the crossover point of the fuzzy set $R$ and the
slope at the crossover point. The fuzzy set has the maximum
uncertainty at the crossover point and the choice of $c=2.2$ is
motivated by the fact that the two distributions corresponding to the
blue cloud and red sequence merge together at $(u-r)\sim2.2$. The
value of $a$ is chosen so as to ensure that the galaxy with largest
and smallest $(u-r)$ colour respectively have their membership
function $1$ and $0$ in the fuzzy set $R$. Once the fuzzy set for
`redness' is defined, the fuzzy set $B$ for `blueness' can be simply
obtained by taking a fuzzy complement of set $R$. The membership
function $\mu_{B}(u-r)$ of the
fuzzy set $B$ is defined as,
\begin{eqnarray}
\mu_{B}(u-r)=1-\mu_{R}(u-r), \,  \forall (u-r) \in X
\label{eq:fuzzy4}
\end{eqnarray}
The fuzzy set $G$ for `greenness' can be defined by taking a fuzzy
intersection of $R$ and $B$. The membership function
$\mu_{G}(u-r)$ of the fuzzy set
$G$ is defined as,
\begin{eqnarray}
\mu_{G}(u-r)= 2 \, min\big{\{}\,
\mu_{R}(u-r),\,
\mu_{B}(u-r)\,\big{\}},\, \forall
(u-r) \in X
\label{eq:fuzzy5}
\end{eqnarray}
Here $min$ denotes the minimum operator. The fuzzy set $G$ has the
maximum height of $0.5$ at the crossover point. The multiplication
with a factor 2 ensure that the galaxies with colour $(u-r)=2.2$ are
maximally green with a membership function of 1
(\autoref{fig:fuzzycol}).

We classify those galaxies as green for which $\mu_{G}$ dominates
$\mu_{B}$ and $\mu_{R}$. The red and blue galaxies are also defined in
a similar manner. In the present analysis, we find that the galaxies
with $2.067<(u-r)<2.33$ are green, $(u-r)\le 2.067$ are blue and
$(u-r)\ge 2.33$ are red (\autoref{fig:fuzzycol}). These cuts provide
us with 45700 red, 34662 blue and 12825 green galaxies.

To study the detailed morphology of green galaxies, we further cross
match their $specObjid$ using the $zoo2MainSpecz$ table derived from
Galaxy Zoo 2 \citep{willett}. We obtain the information regarding the
presence of bulge, bar, disturbed, irregular or merger features in the
green galaxies in our sample. We only consider the classifications
with a debiased vote fraction greater than $0.8$ in each case. The
cross matching yields the required information for 8931 out of 12825
green galaxies.

\subsection{Quantifying the environment of galaxies with the local dimension}

The galaxies reside in different types of geometric environments in
the cosmic web. We quantify the geometric environment of a galaxy
using the local dimension \citep{sarkar09}. The local dimension of a
galaxy is simply based on the number counts of galaxies within a
sphere of radius $R$ centered around it. The number count $N(<R)$ is
expected to scale as,
\begin{eqnarray}
 N(<R)= A R^{D}
\label{eq:ld}
\end{eqnarray}
where $D$ is the local dimension and A is a constant. We vary the
radius of the measuring sphere over a range of length scales $R_1
\hmpc \leq R \leq R_2 \hmpc$. All the galaxies for which the radius of
the measuring sphere can be varied in this range and there are at
least 10 galaxies within the two concentric spheres of radius $R_1$
and $R_2$, are included in the analysis. For each valid centers, we
fit the measured number counts $N(<R)$ within $R_1$ and $R_2$ to
\autoref{eq:ld}. The best fit values of A and $D$ are determined using
a least-square fitting. We also estimate the associated $\chi^2$ value
using the fitted and observed values of $N(<R)$. We further restrict
our analysis to only those centers for which chi-square per degree of
freedom $\frac{\chi^2}{\nu} \leq 0.5$ \citep{sarkar19}. We consider
$R_1=2 \hmpc$ throughout the present analysis. We have chosen $R_2=10
\hmpc$ and $R_2=40 \hmpc$ to probe the geometric environments of
galaxies on two different length scales. The geometric environment
around a galaxy within length scale range $R_1 \hmpc \leq R \leq R_2
\hmpc$ is characterized by its local dimension $D$. The galaxies
residing at the center of a straight filament are expected to have a
local dimension of $D=1$.  A local dimension of $D=2$ would represent
the galaxies lying within a sheet-like structure whereas the galaxies
with $D=3$ are expected to be distributed homogeneously in a three
dimensional volume. The cosmic web is an interconnected complex
network of different morphological components which vary widely in
their shapes and sizes. Very often, the measuring sphere may include
multiple types of morphological components (e.g. a filament and a
sheet) leading to intermediate values of local dimension $D$. We
classify the galaxies belonging to different types of geometric
environments by assigning a specific range of local dimension to each
class (\autoref{tab:gclass}). In this classification, the $D1$-type
galaxies are either located near the center of straight filaments or
they are part of short tendrils of galaxies which links filaments and
penetrates into voids \citep{alpaslan14}. The $D2$-type galaxies
inhabit sheet-like environments and the $D3$-type galaxies mostly
belong to fields. In a recent work, Pandey \& Sarkar \citep{pandey20a}
explored the fraction of red and blue galaxies in different
environments and reported a higher fraction of red galaxies in
filaments than sheets at all luminosities. In the same work, they show
that the galaxies with $D1$-type environment mostly resides in
straight filaments whereas those with $D2$ and $D3$-type environments
are primarily associated with the sheetlike and field environments.
The distributions of $D1$, $D2$ and $D3$-type galaxies in the SDSS
were shown in Figure 3 of that paper. The galaxies in the $D1.5$ class
represent the geometric environment which are intermediate between
filaments and sheets. Similarly the galaxies of $D2.5$-type populate
the intervening regions between sheets and fields.The local dimension
of a galaxy would also depend on the length scales considered in the
analysis. Keeping this in mind, we have carried out our analysis on
two different length scale ranges.

We identify the green valley population using the fuzzy set theory
method described in Section 2. We study the abundance and different
properties of the green galaxies in different geometric environments
of the cosmic web. We estimate the errorbars from 10 jackknife samples
prepared from the data. The jackknife samples are prepared by randomly
discarding $25\%$ galaxies from the original volume limited sample.

It may be noted that the local dimension can not be calculated for all
the galaxies in our volume limited sample. Also the number of galaxies
for which the local dimension can be calculated would decrease with
the increasing length scales. We find that the local dimension for
36031 red, 27482 blue and 10136 green galaxies can be computed at
$R_2=10 \hmpc$ whereas at $40 \hmpc$, the local dimension can be
computed for 14441 red, 11591 blue and 4207 green galaxies.

\subsection{Testing the differences in stellar mass, SFR and stellar age of green galaxies in different environments with the Kolmogorov-Smirnov test}

If environment plays a significant role in the transition of green
valley galaxies then the distributions of their properties should not
be same in different environments. For any given property of green
galaxy, we compare the cumulative distributions in two different
environments with the two-sample Kolmogorov-Smirnov (KS) test. The KS
test does not make any assumptions about the distributions. The null
hypothesis associated with the test assumes that the chosen property
of the green galaxies from different environments are actually sampled
from identical distributions. We define the supremum difference
$D_{KS}$ between the two cumulative distribution functions as,
\begin{eqnarray}
D_{KS} & = & \sup_{X} \, \, \{ \,\, | f_{1,
  m}(X)-f_{2,m}(X) | \,\, \}
  \label{eqn:Dks}
\end{eqnarray}
where $f_{1,m}(X)$ and $f_{2,m}(X)$ are the cumulative distribution
functions of the chosen property ($X$) of the green galaxies at the
$m^{th}$ bin in type-1 and type-2 environments respectively. Here the
properties that we have chosen are the stellar mass, star formation
rate and stellar age. The type-1 and type-2 can be any environments
out of $D1$, $D1.5$, $D2$, $D2.5$ and $D3$. We have $m \in
\{1,2,3...., N^{'} \}$ and $\sup$ denotes the supremum of all the
($N_{1}^{'}+N_{2}^{'}$) differences such that
$\sum_{m=1}^{N^{'}}{f_{1, m}(X)}=\sum_{m=1}^{N^{'}}{f_{2, m}(X)}=1$.

The critical value of the supremum difference associated with a given
significance level ($\alpha$) is given by,
 \begin{eqnarray}
D_{KS} (\alpha) & = & \sqrt{- \ln \left( \frac{\alpha}{2} \right) \,\,
  \, \frac{ N_{1}^{'} + N_{2}^{'}}{2 N_{1}^{'} N_{2}^{'}}}
\label{eqn:aks}
\end{eqnarray}
Here $N_{1}^{'}$ and $N_{2}^{'}$ are the number of green galaxies in
type-1 and type-2 environments respectively. When $D_{KS}>D_{KS}
(\alpha)$, we can reject the null hypothesis at a significance level
$\alpha$. The null hypothesis can be tested at different significance
level to find if the distribution of a given property of green
galaxies are significantly different in the two separate environments
considered.

\begin{table*}
\centering
\begin{tabular}{|c|c|}
\hline
Local dimension & Geometric environment \\
\hline
$ 0.75 \le D < 1.25 $ & $D1$ \\
$1.25 \le D < 1.75$  & $D1.5$ \\
$ 1.75 \le D < 2.25$ & $D2$ \\
$ 2.25 \le D < 2.75 $ &  $D2.5$ \\
$ D \ge 2.75 $ & $D3$ \\

\hline
\end{tabular}
\caption{This table defines the different geometric environments in
  the cosmic web based on the local dimension of a galaxy.}
\label{tab:gclass}
\end{table*}

\section{Results}
\subsection{Fractions of red, blue and green galaxies in different environments}
\begin{figure*}[htbp!]
\centering
\includegraphics[width=7cm]{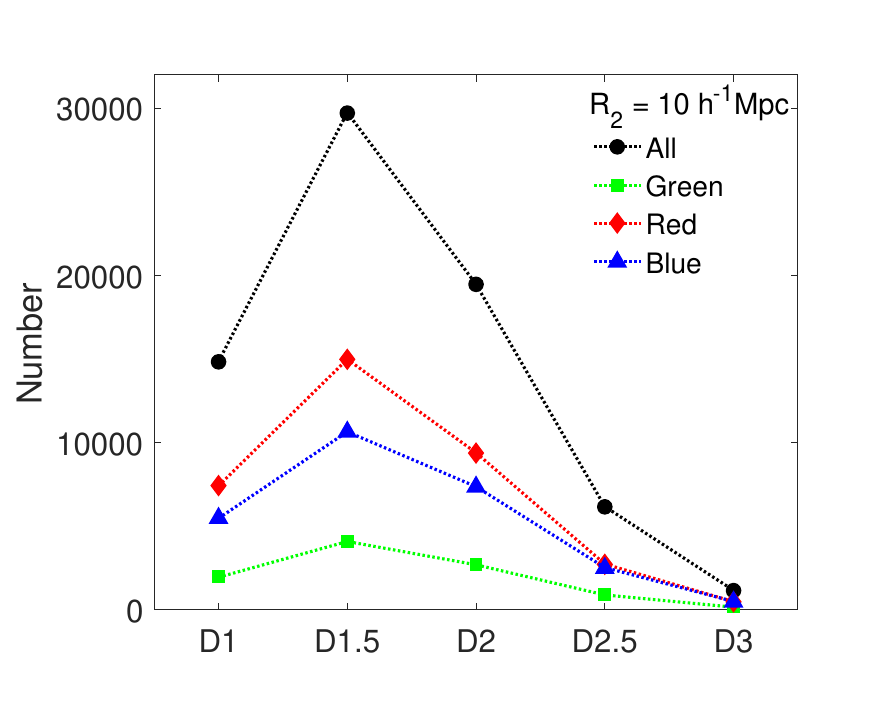}\hspace{0.1cm} 
\vspace{-0.6cm}
\includegraphics[width=7cm]{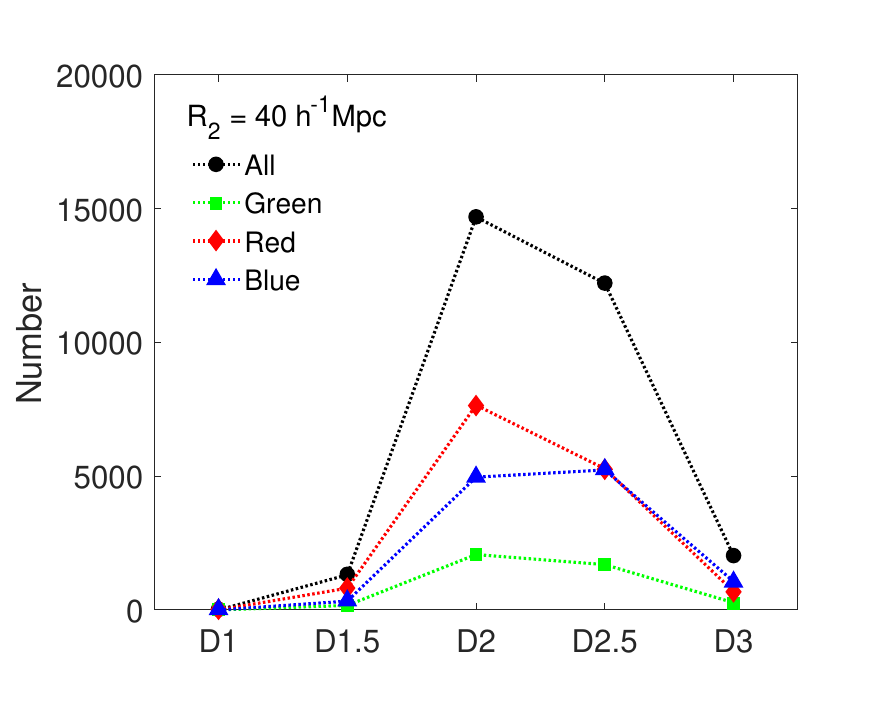}
\includegraphics[width=7cm]{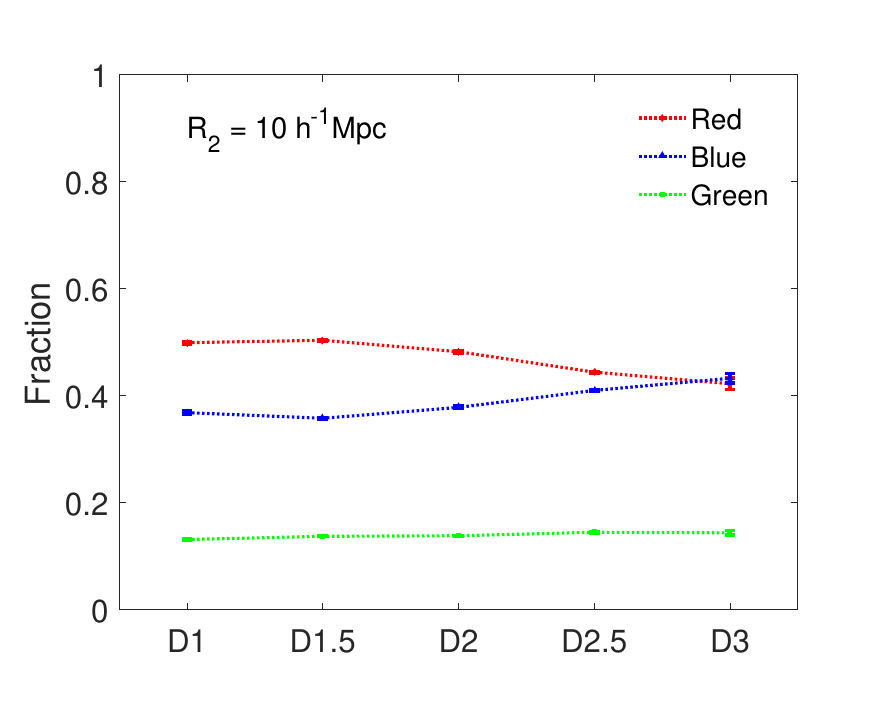}\hspace{0.1cm}
\includegraphics[width=7cm]{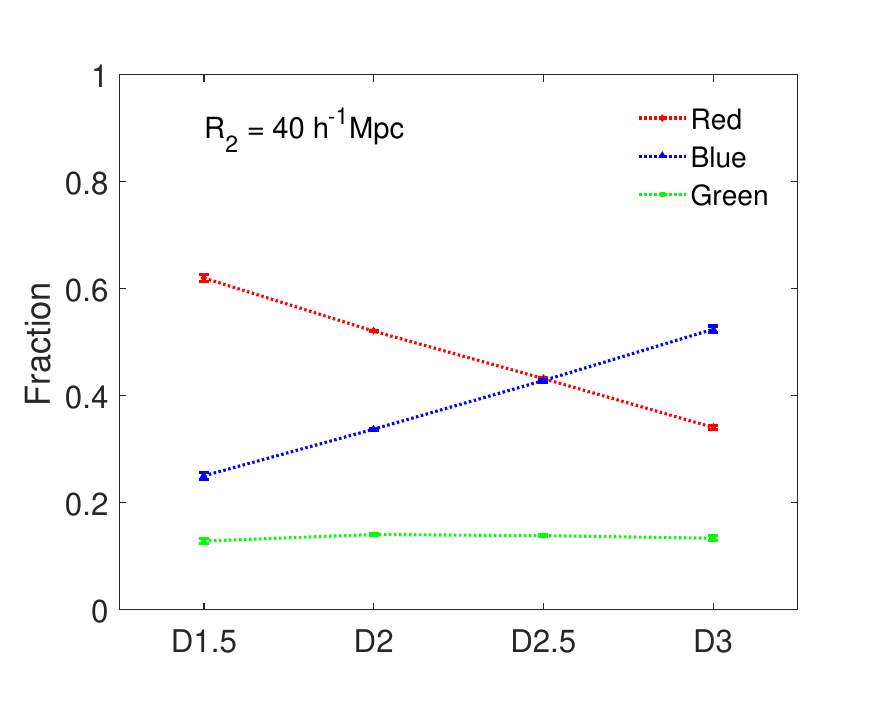}
\caption{The top left and top right panel shows the number of red,
  blue and green galaxies in different environments for $R_2=10 \hmpc$
  and $R_2=40 \hmpc$. The respective fractions are shown in the bottom
  left and bottom right panels. The $1-\sigma$ errorbars shown here are
  estimated using 10 jackknife samples drawn from the volume limited
  sample.}
\label{fig:fgreen1}
\end{figure*}

We show the number of red, blue and green galaxies available in
different environments of the cosmic web in the top two panels of
\autoref{fig:fgreen1}. The top left and right panel of this figure
respectively correspond to $R_2=10 \hmpc$ and $R_2=40 \hmpc$. The top
left panel shows that the number of galaxies for each class peaks at
local dimension of $D=1.5$. This implies that the most abundant
structures on length scales of $10 \hmpc$ have a morphology which is
intermediate between filaments ($D=1$) and sheets ($D=2$). These
structures could represent the curved filaments or the intersecting
regions between filaments and sheets. The number of galaxies in
different classes peak at the local dimension of $D=2$ in the top
right panel which indicates that the cosmic web is dominated by the
sheet-like patterns when analyzed on a length scale of $40
\hmpc$. There is a clear shift of the location of the peak with
increasing length scales which suggests that there is a gradual
transition of the local dimension of the morphological patterns in the
cosmic web with increasing length scales. The fraction of red, blue
and green galaxies in different environments for $R_2=10 \hmpc$ and
$R_2=40 \hmpc$ are respectively shown in the two bottom panels of
\autoref{fig:fgreen1}. The results suggest that the fraction of red
galaxies decreases with the increasing local dimension of their host
environment. We see a reverse trend for the blue galaxies. These
trends were reported earlier in a paper by \citep{pandey20a}. Here, we
would like to make it clear that these results in
\autoref{fig:fgreen1} are shown here only for the sake of
completeness. In the present work, we are primarily interested in the
green valley galaxies. We would like to understand the relative
importance of internal physical processes and external environments in
driving the transition in green valley galaxies.

We also show the fraction of green galaxies in different geometric
environments of the cosmic web in the two bottom panels of
\autoref{fig:fgreen1}. We find that $\sim 15\%$ galaxies in each
geometric environment are green. The fraction of green galaxies are
nearly independent of the local dimension of their host environments.

Mass of a galaxy plays an important role in quenching star
formation. The star formation is quenched in galaxies above a critical
mass. The massive galaxies are more common in high density
environments. It is known that the geometric environments with smaller
local dimension have higher local densities \citep{pandey20a}. In
general, filaments are denser than sheets and the sheets are denser
than the fields. So a higher fraction of red galaxies in filaments
than sheets and fields may simply arise due to a higher abundance of
more massive galaxies in filaments compared to the other geometric
environments. A similar argument applies for the blue galaxies which
have a relatively lower mass and are more abundant in low density
environments with higher local dimension. Thus a higher fraction of
galaxies are expected to be mass quenched in environments with smaller
local dimension. The green galaxies are intermediate between red and
blue galaxies and it is interesting to see that the fraction of green
galaxies is nearly independent of environment in
\autoref{fig:fgreen1}.

\begin{figure*}[htbp!]
\centering
\includegraphics[width=7cm]{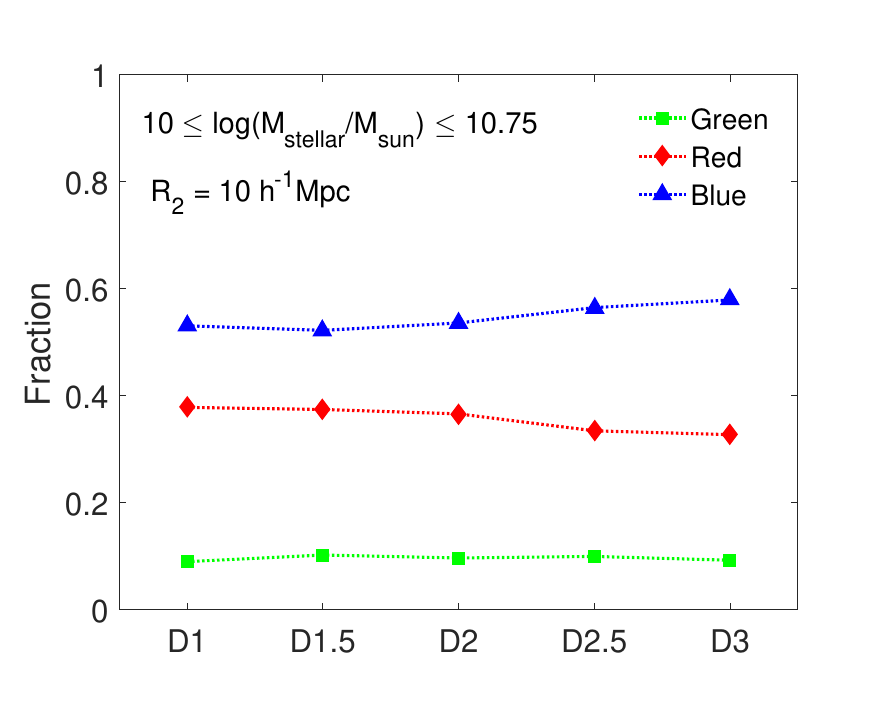}\hspace{0.1cm}
\vspace{-0.6cm}
\includegraphics[width=7cm]{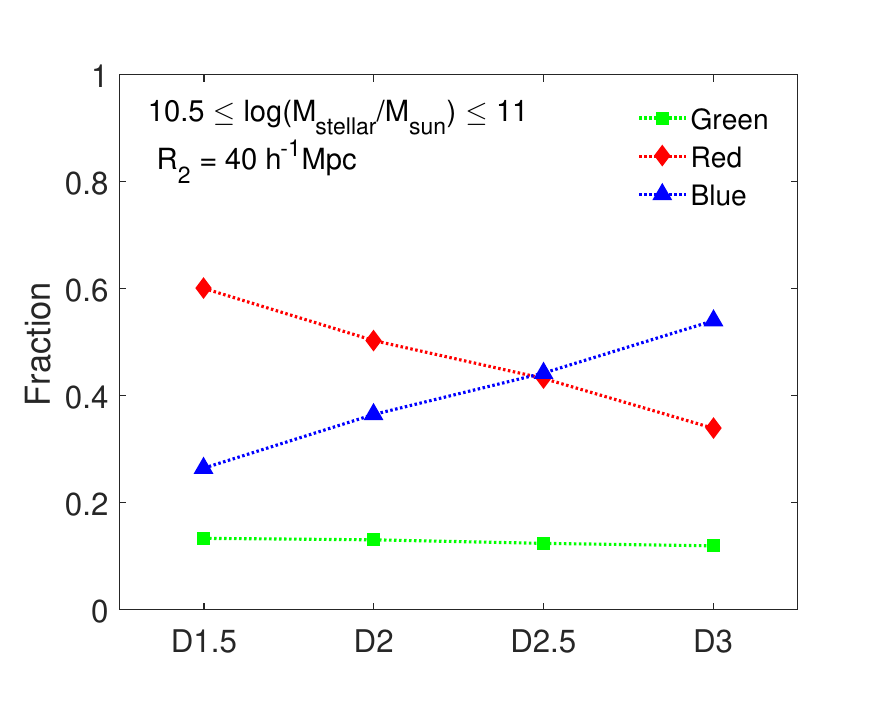}
\includegraphics[width=7cm]{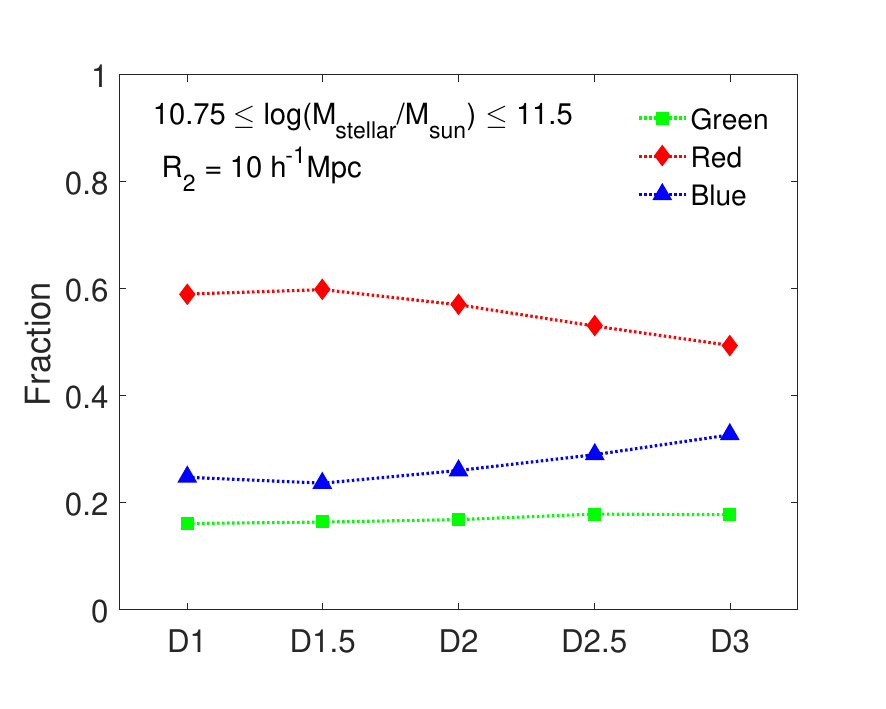}\hspace{0.1cm}
\includegraphics[width=7cm]{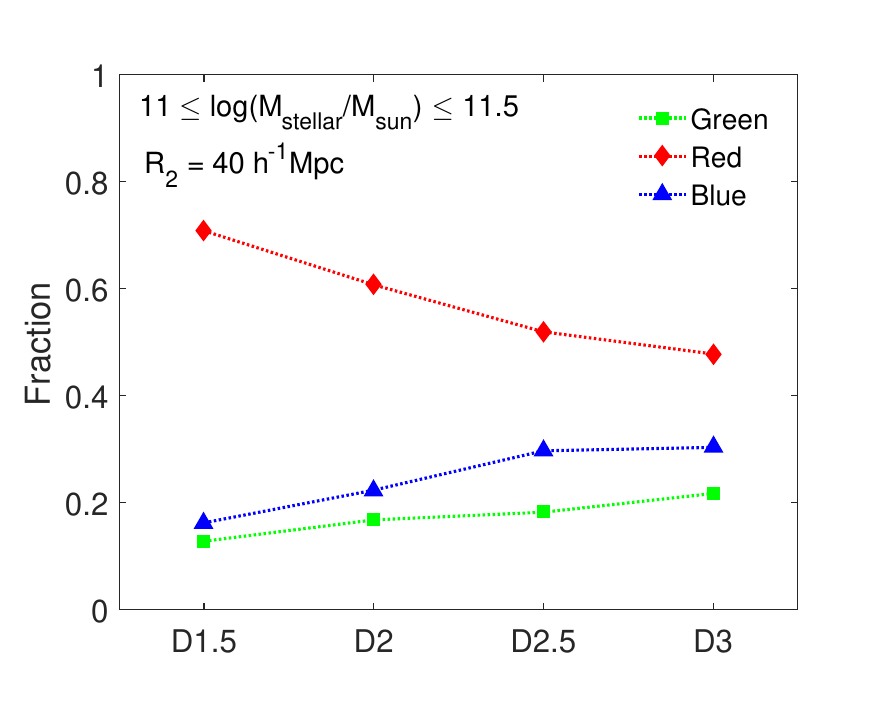}
\caption{The top left and bottom left panels show the fractions of
  red, blue and green galaxies in different environments for $R_2=10
  \hmpc$. These correspond to two different mass bins as mentioned in
  the panels. The results for two different mass bins for $R_2=40
  \hmpc$ are shown in the top right and bottom right panels of the
  figure. We estimate the $1-\sigma$ errorbars using 10 jackknife
  samples drawn from the volume limited sample. The size of the
  error-bars are smaller compared to the size of the symbols used in
  each panel.}
\label{fig:fgreen2}
\end{figure*}

\begin{figure*}[htbp!]
\centering
\includegraphics[width=7cm]{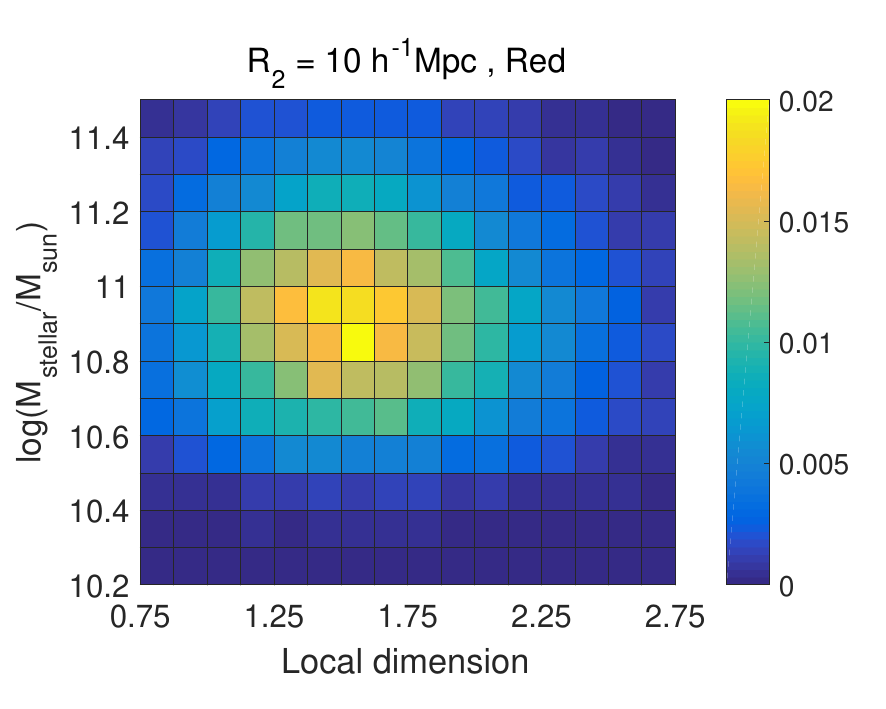}\hspace{0.1cm}
\vspace{-0.49cm}
\includegraphics[width=7cm]{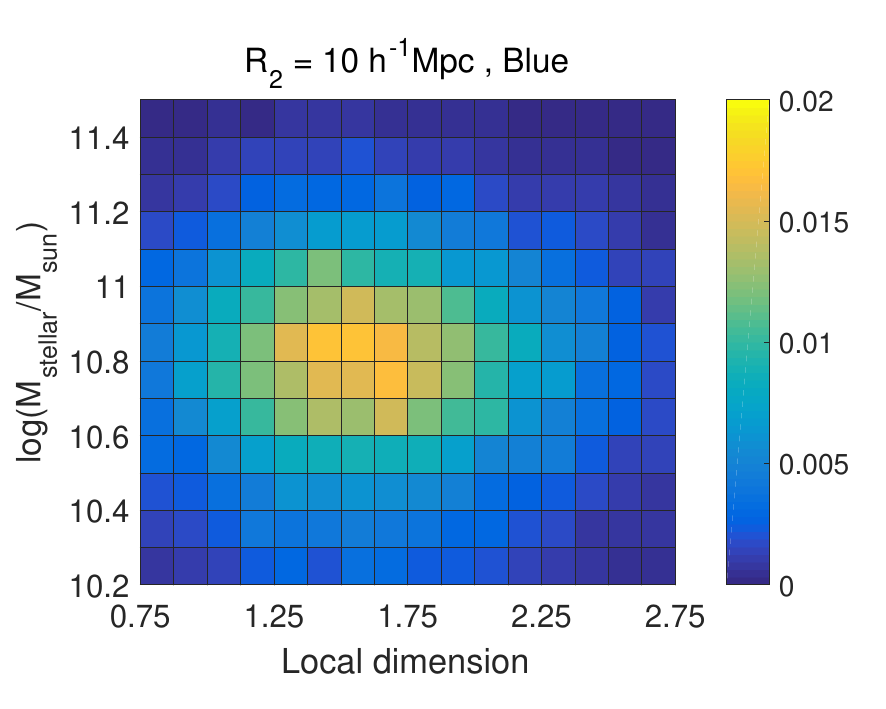}\hspace{0.1cm}
\includegraphics[width=7cm]{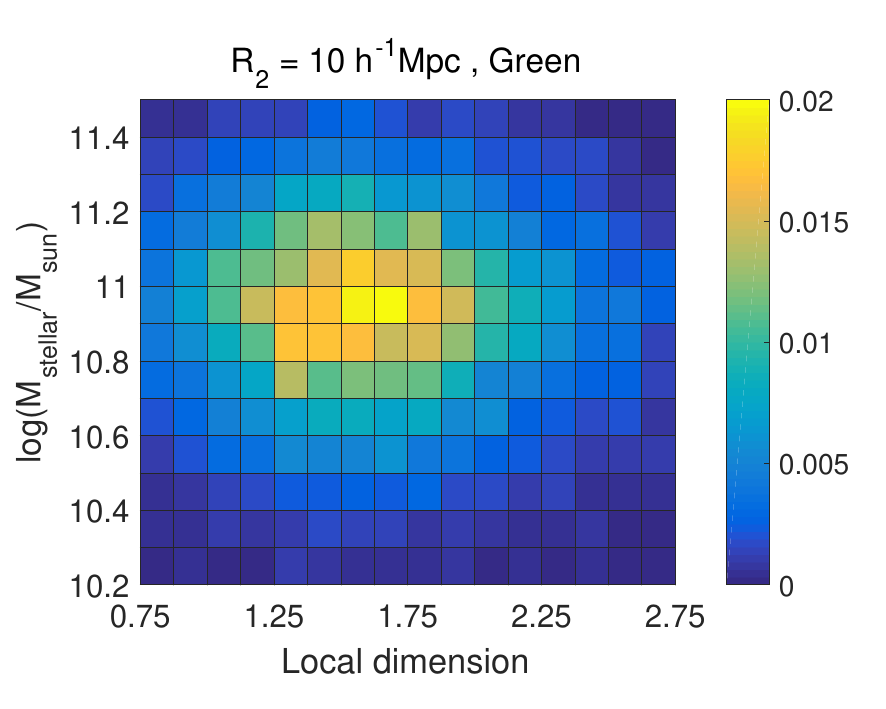}\hspace{0.1cm}
\includegraphics[width=7cm]{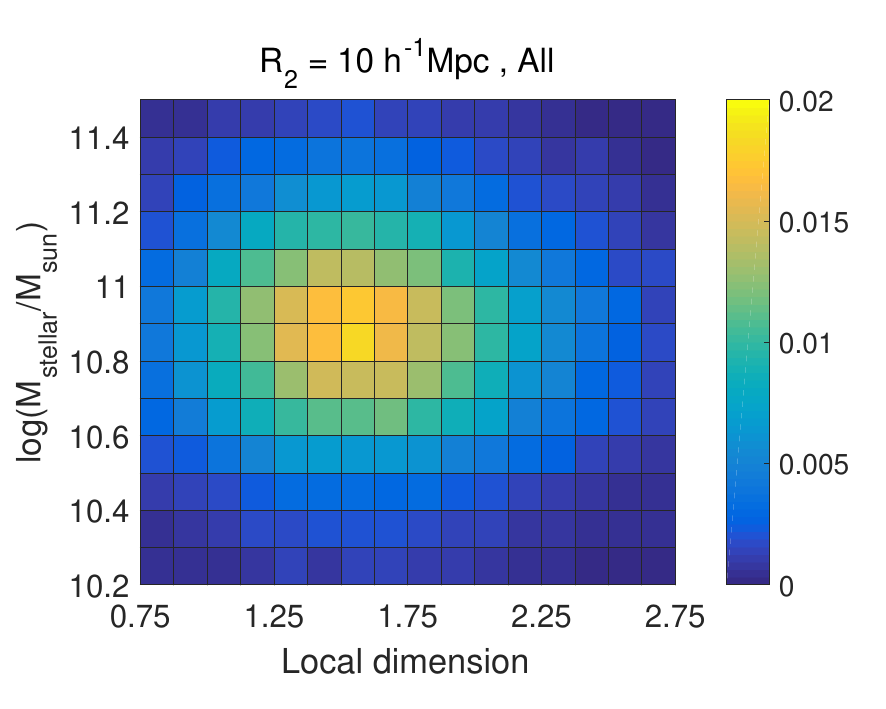}\hspace{0.1cm}
\caption{The above plot shows the fraction of
    different types of galaxies in different regions of the stellar
    mass-local dimension plane. Here we have used the actual values of
    the local dimension for the galaxies without applying the
    classification scheme listed in \autoref{tab:gclass}. The top two
    panels of this figure show the results for the red and blue
    galaxies. The results for the green galaxies and the combined
    sample are shown in the bottom two panels. The colour bars
    represent the fraction of galaxies.}
\label{fig:hist2d}
\end{figure*}

\begin{figure*}[htbp!]
\centering
\includegraphics[width=7cm]{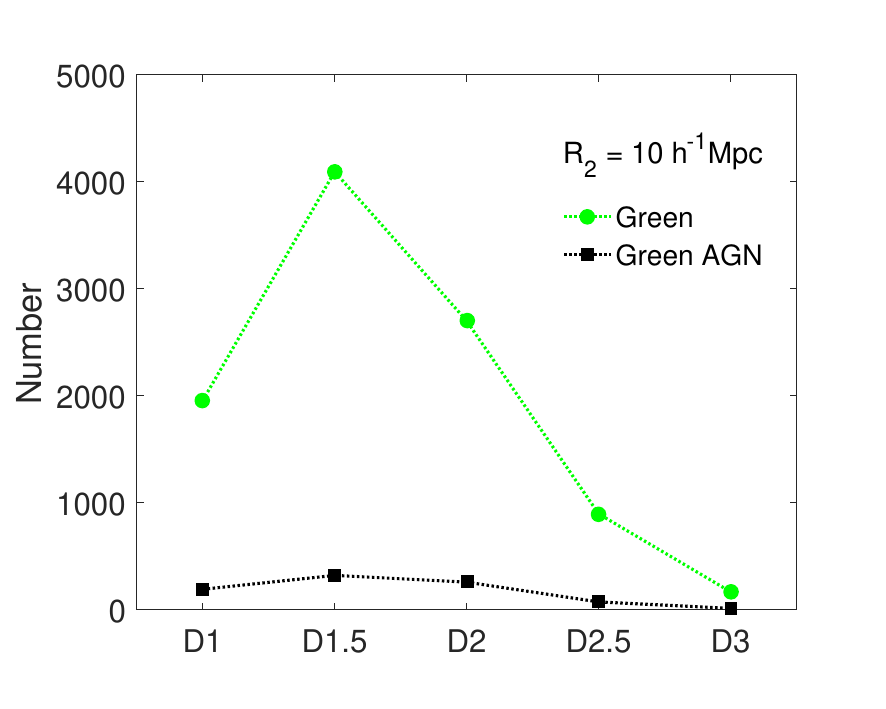}\hspace{0.1cm}
\vspace{-0.6cm}
\includegraphics[width=7cm]{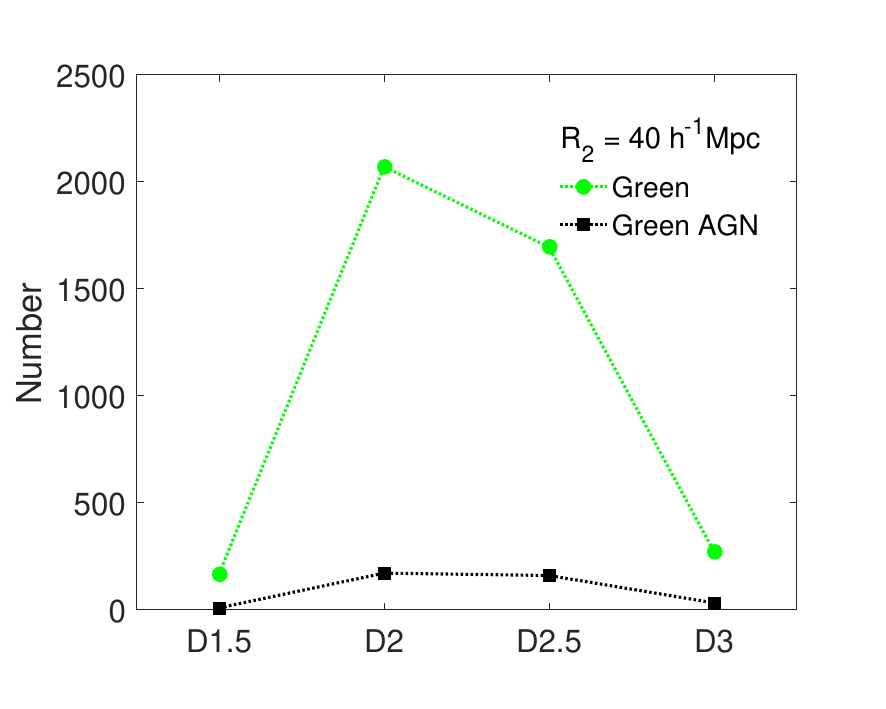}\\
\includegraphics[width=7cm]{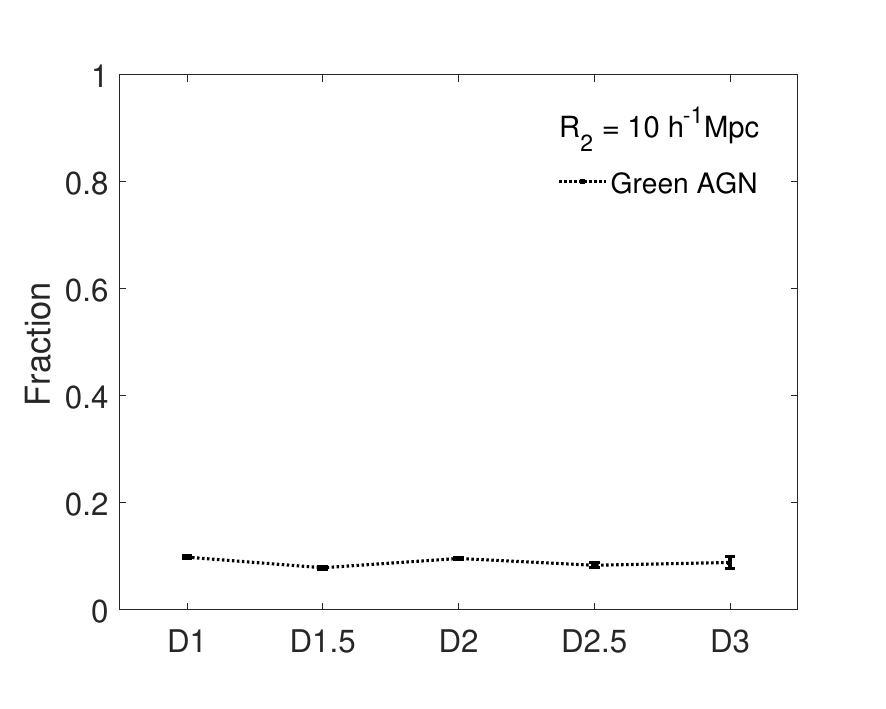}\hspace{0.1cm}
\includegraphics[width=7cm]{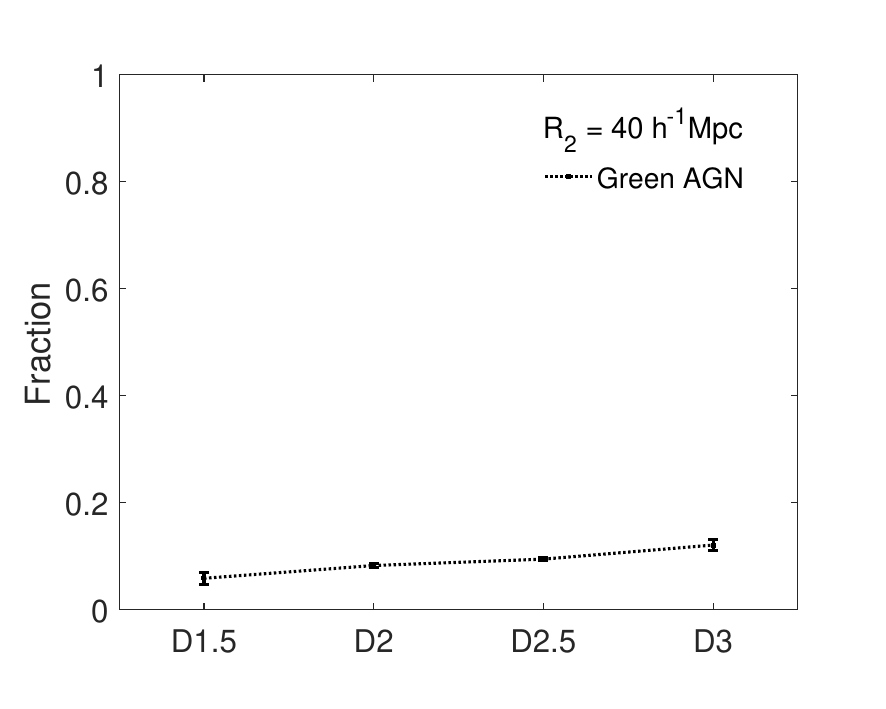}\\
\caption{The top left and top right panels show the number of green
  galaxies and green AGNs in different environments for $R_2=10 \hmpc$
  and $R_2=40 \hmpc$ respectively. The bottom left and right panels
  shows the respective fractions of AGNs at each environment. The
  $1-\sigma$ errorbars shown are estimated using 10 jackknife samples
  drawn from the green galaxy sample.}

\label{fig:agn}
\end{figure*}

\begin{figure*}[htbp!]
\centering
\includegraphics[width=7cm]{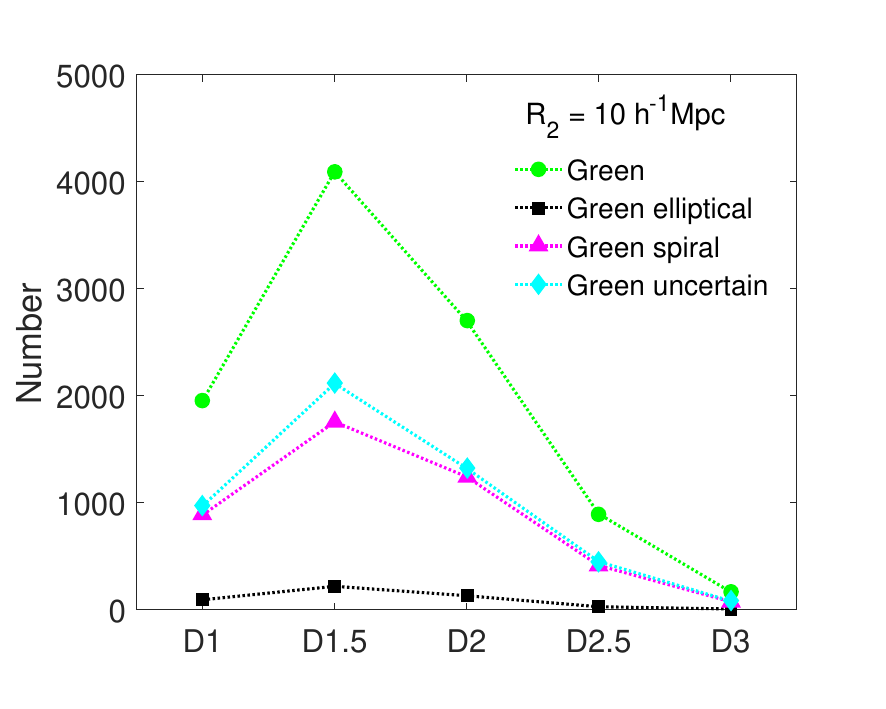}\hspace{0.1cm}
\vspace{-0.6cm}
\includegraphics[width=7cm]{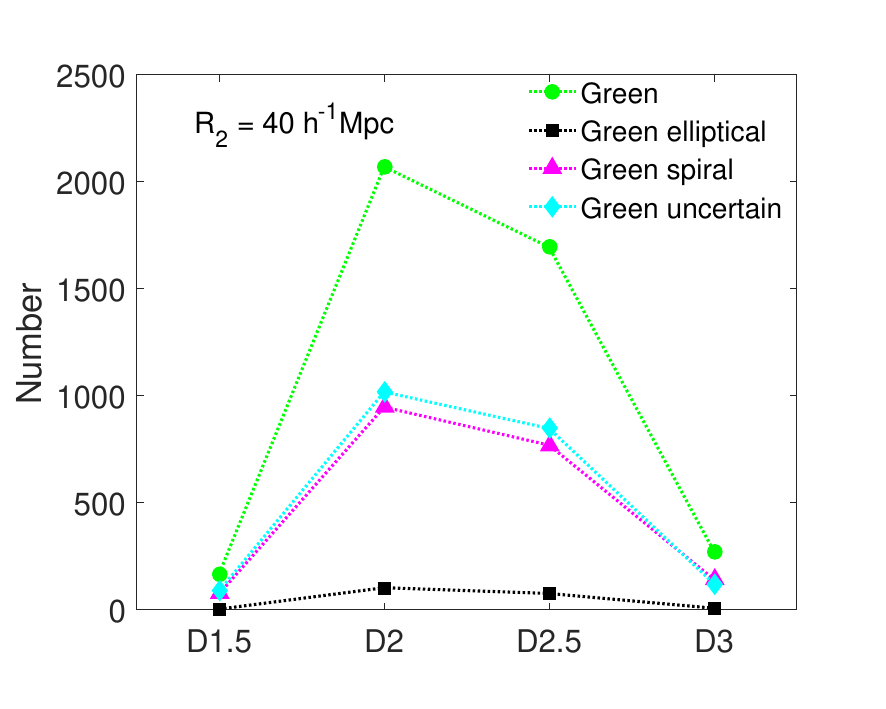}
\includegraphics[width=7cm]{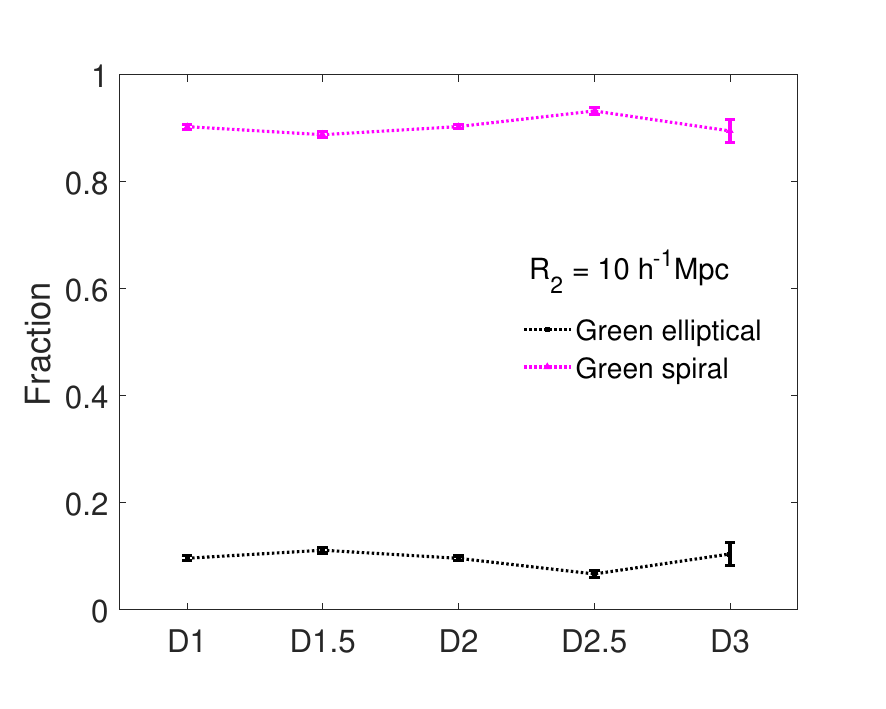}\hspace{0.1cm}
\includegraphics[width=7cm]{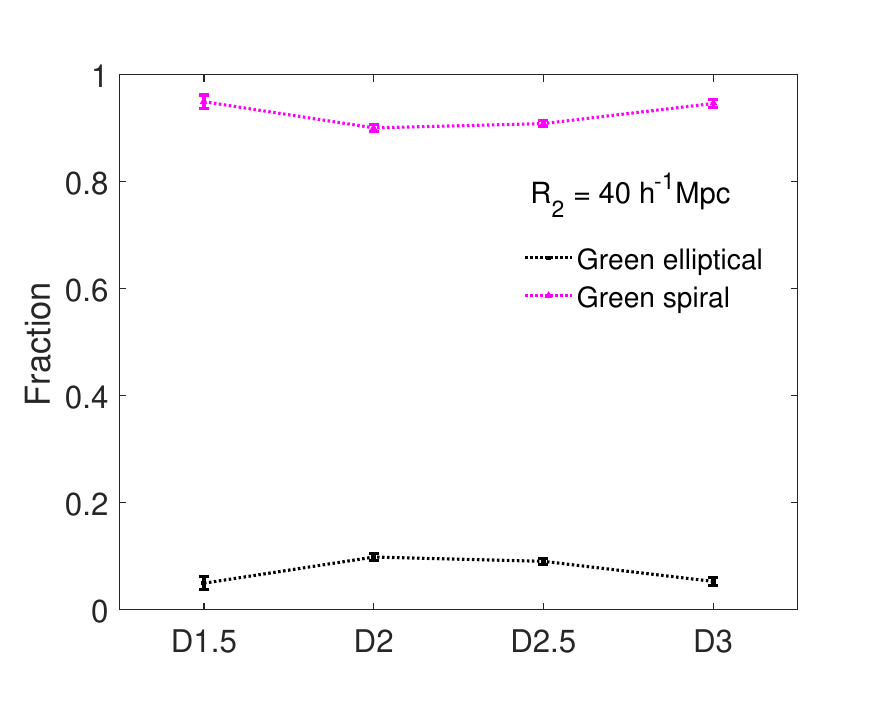}
\caption{The top left and top right panel show the number of green
  galaxies with elliptical, spiral and uncertain morphology in
  different environments for $R_2=10 \hmpc$ and $R_2=40 \hmpc$
  respectively. The bottom left and bottom right panels show the
  respective fractions of classified green galaxies with spiral and
  elliptical morphology. The $1-\sigma$ errorbars shown are estimated
  using 10 jackknife samples drawn from the sample of green galaxies.}

\label{fig:morph1}
\end{figure*}

\begin{figure*}[htbp!]
  \centering
\includegraphics[width=7cm]{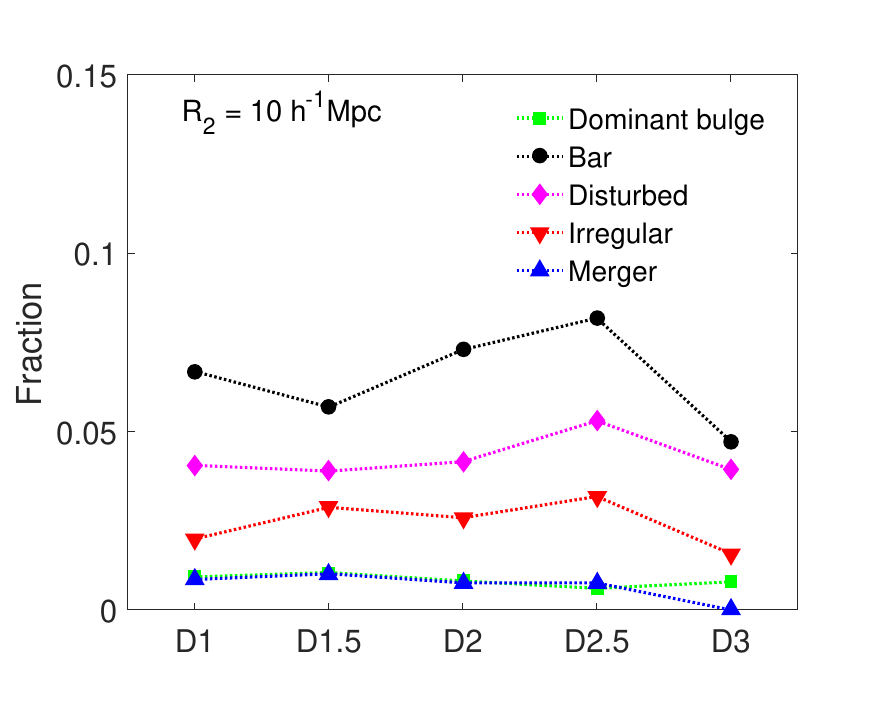}\hspace{0.1cm}
\includegraphics[width=7cm]{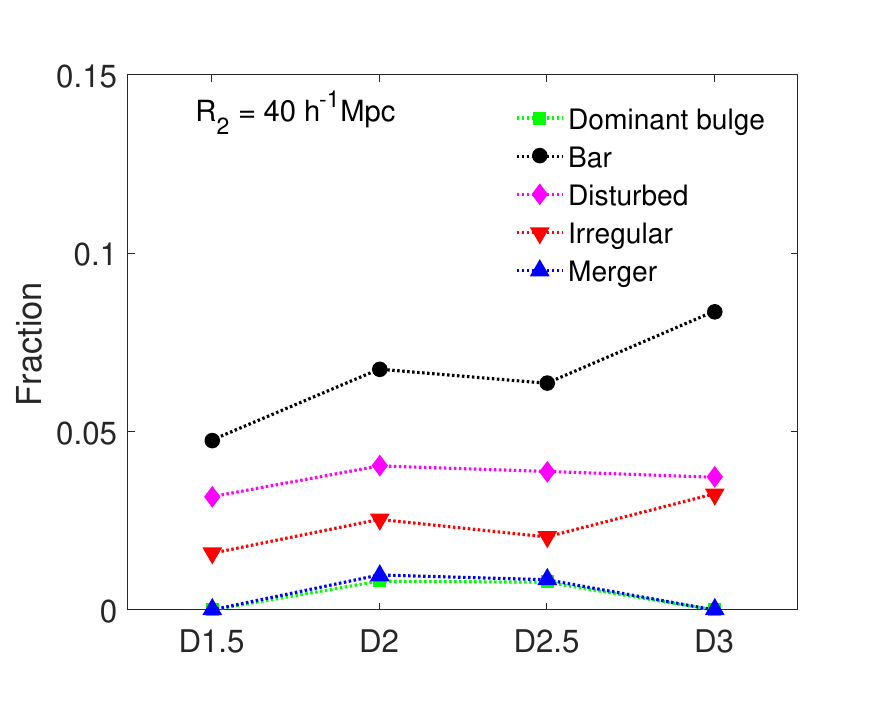}
\caption{The left and right panels separately show the fraction of
  green galaxies with the dominant bulge, bar and those with the
  disturbed, irregular and merger features in different environments
  for $R_2=10 \hmpc$ and $R_2=40 \hmpc$.}
\label{fig:morph2}
\end{figure*}

\subsection{Mass dependence of the fractions of red, blue and green galaxies in different environments}

In order to understand the role of mass quenching in different
environments, we divide the red, blue and green galaxies into two
different mass bins and study their fractions in each environment.
The top left and bottom left panels of \autoref{fig:fgreen2} show the
results in two different mass bins for $R_2=10 \hmpc$. The top left
panel of \autoref{fig:fgreen2} shows that at lower masses, a higher
fraction of blue galaxies are present at each environment. This is
opposite to what is observed in \autoref{fig:fgreen1}. Here the
fraction of blue galaxies also increases with increasing local
dimension as noted earlier in \autoref{fig:fgreen1}. The fraction of
red galaxies are smaller than blue galaxies at each environment and
the red fraction decreases with the increasing local dimension as
before.  Only $\sim 10\%$ galaxies at each environment are green and
the fraction of green galaxies is independent of environment in the
lower mass bin.  Now we shift our attention to the bottom left panel
where the fractions are shown for the galaxies in the higher mass
bin. We note that at higher mass bin, each environment is dominated by
the red galaxies. The fractions of red galaxies are now significantly
higher ($2-3$ times) compared to blue galaxies at each
environment. The fraction of green galaxies ($\sim 18\%$ in this case)
also increases in the higher mass bin which again remains nearly
independent of environment. We observe a change in the green fraction
in this case due to the fact that all the galaxies in the higher mass
bin have masses above the critical mass ($3 \times
10^{10}\,M_{\odot}$) required for mass quenching whereas a significant
number of galaxies in the lower mass bin have masses below this
critical value. The result suggest that mass quenching may play an
important role in suppressing star formation in the green valley
galaxies.

The corresponding results for $R_2=40 \hmpc$ are shown in the top
right and bottom right panels of \autoref{fig:fgreen2}. We see that
the red galaxies dominate each environment in both the mass
bins. Noticeably at higher masses, the fraction of blue galaxies show
a decrease in each environment. The decrease is most pronounced for
the field galaxies ($D3$-type). This implies that there are more
galaxies yet to be mass quenched in the low density regions as
compared to the high density environments. Increase in red fraction
and decrease in blue fraction in all environments suggest that the
galaxies with higher masses are equally likely to be quenched
irrespective of their environment. Once again, we observe that the
fraction of green galaxies remains within $15\%-20 \%$ in all the
environments. Only a mild increase in the fraction of green galaxies
are observed in the higher mass bin. It may be noted that the stellar
mass of each galaxies in both the mass bins are greater than the
critical mass.

Our analysis shows that the fraction of already quenched red galaxies
depends on environment at fixed stellar mass, and also depends on
stellar mass at fixed environment (\autoref{fig:fgreen2}). The
sensitivity of the red fractions to the stellar mass and the
environment suggests that besides the mass quenching, the external
influences of the environment may also play a significant role in
quenching the star formation in these galaxies \citep{peng}. The green
valley galaxies are in the process of being quenched and in principle
both the environment and mass are capable of quenching them. But the
top left and bottom left panels of \autoref{fig:fgreen2} show that at
a fixed stellar mass, the fraction of green galaxies is nearly
independent of environment whereas at a fixed environment the fraction
increases from $10\%$ to $20\%$ in any given environment. This
indicates that quenching in the green valley galaxies are largely
driven by their mass and other internal physical properties rather
than their external environment.

  We also show the fraction of different types of
  galaxies present in different regions of the local dimension-stellar
  mass plane in \autoref{fig:hist2d}. The top left, top right and
  bottom left panels of this figure show the results for the red, blue
  and green galaxies respectively. The results show that the red, blue
  and green galaxies reside in all types of environments. However they
  preferentially occupy the environments with local dimension $\sim
  1.5$ when the environments are examined on a length scales of $10
  \hmpc$. Such a trend is already seen in \autoref{fig:fgreen1}. The
  top two panels of \autoref{fig:hist2d} show that the red galaxies
  show a shift towards the higher stellar masses compared to the blue
  galaxies in the local dimension-stellar mass plane. This suggests a
  role of stellar mass in quenching the red galaxies in all types of
  environments. Interestingly, the green galaxies occupy a similar
  region as inhabited by the red galaxies in the local
  dimension-stellar mass plane. The strong environmental dependence of
  the red fraction and the environmental independence of the green
  fraction in \autoref{fig:fgreen2} suggest that the red and green
  galaxies may have acquired their higher masses through different
  routes.

\subsection{AGN activity in green galaxies in different environments}

AGN feedback is believed to play an important role in quenching star
formation in galaxies. It prevents cooling of hot halo gas and
supports long term shutdown of star formation in massive galaxies. We
explore the fraction of AGN in green galaxies in different
environments of the cosmic web. The top left and right panels of
\autoref{fig:agn} show the number of green galaxies and green AGN in
different environments for $R_2=10 \hmpc$ and $R_2=40 \hmpc$
respectively. The fractions of green AGNs in different environments
for these two length scales are shown in the two bottom panels of
\autoref{fig:agn}. We find that $\sim 10\%$ green galaxies host AGN in
each environment. The fraction of AGN in green galaxies are
independent of environments and the associated length scales. This
indicates that the AGN activity in green galaxies could be a result of
secular evolution. At each environment, only a small fraction of green
galaxies host AGN. The destruction of the gas reservoirs in galaxies
by AGN-driven winds provide an effective channel for
quenching. However, the transition in majority of green galaxies can
not be explained by the AGN driven quenching of star formation.


\subsection{Morphology of green galaxies in different environments}

The morphology and the star formation activity of a galaxy are known
to be closely correlated with each other. In order to understand the
role of morphological quenching in the green valley galaxies, we study
their morphology as classified visually by a large number of
volunteers in the Galaxy Zoo \citep{lintott08} and the Galaxy Zoo 2
project \citep{willett}. We find that out of 12825 green galaxies
there are 640 ellipticals , 5655 spirals and 6530 galaxies with
uncertain morphology. We first show the number of spirals, ellipticals
and galaxies with uncertain morphology in different environments for
$R_2=10 \hmpc$ and $R_2=40 \hmpc$ in the top two panels of
\autoref{fig:morph1}. We plot the respective fractions of spirals and
ellipticals in different environments in the two bottom panels of this
figure. The results show that $\sim 95\%$ classified green galaxies
are spirals and only $\sim 5\%$ galaxies are ellipticals in each
environments. These $5\%$ ellipticals in the green valley could be the
passive galaxies which may have underwent a wet merger causing a
rejuvenation of star formation \citep{kaviraj09,thomas10}. The
majority of the classified green galaxies are spirals. We obtain the
detailed morphology of a subset of the green galaxies (8931 out of
12825) from Galaxy Zoo 2. We find that out of these 12825 green
galaxies, there are 82 galaxies with dominant bulge, 587 galaxies with
bar, 390 galaxies with disturbed features, 231 galaxies with irregular
features and 76 galaxies with merger features. In each case, we only
consider the classifications with a debiased vote fraction greater
than 0.8.  We show the fractions of green galaxies having dominant
bulge, bar and those with disturbed, merger and irregular features in
different environment in the two panels of \autoref{fig:morph2}. In
both the panels, these fractions are nearly independent of
environment.

\begin{figure*}[htbp!]
\centering
\includegraphics[width=7cm]{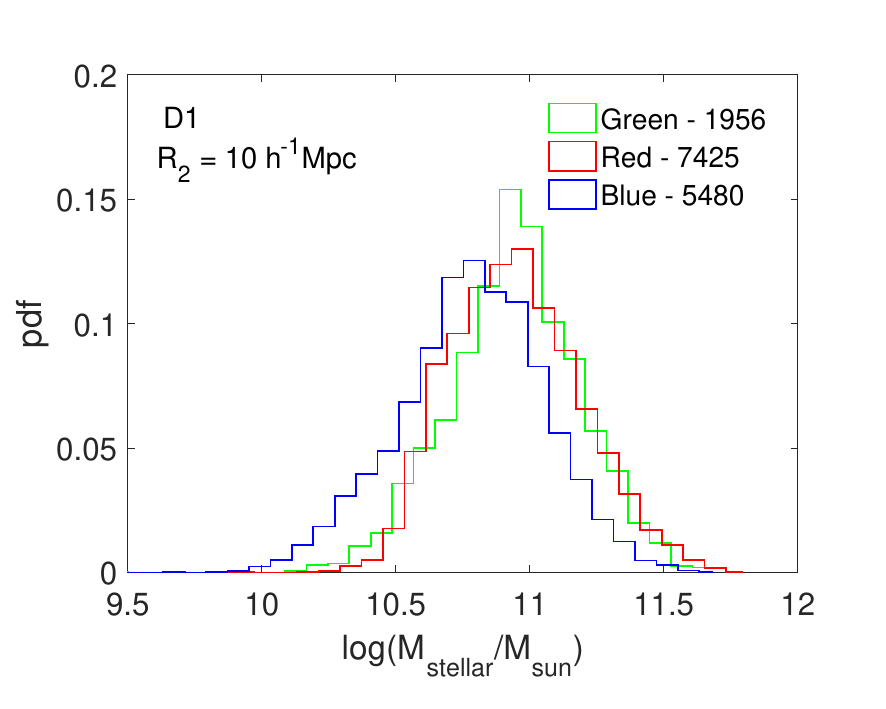}\hspace{0.1cm}
\vspace{-0.48cm}
\includegraphics[width=7cm]{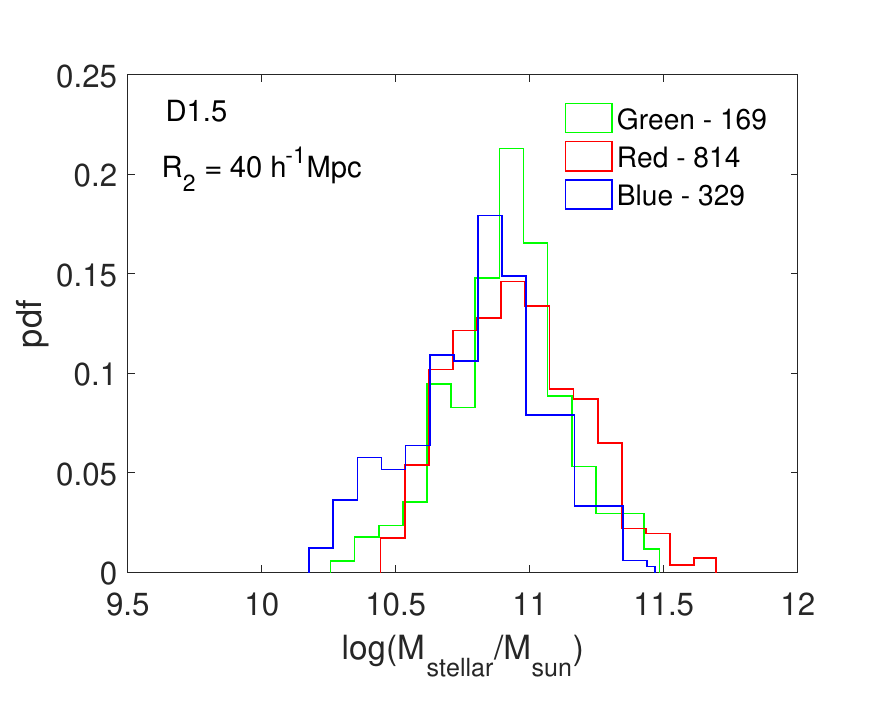}
\includegraphics[width=7cm]{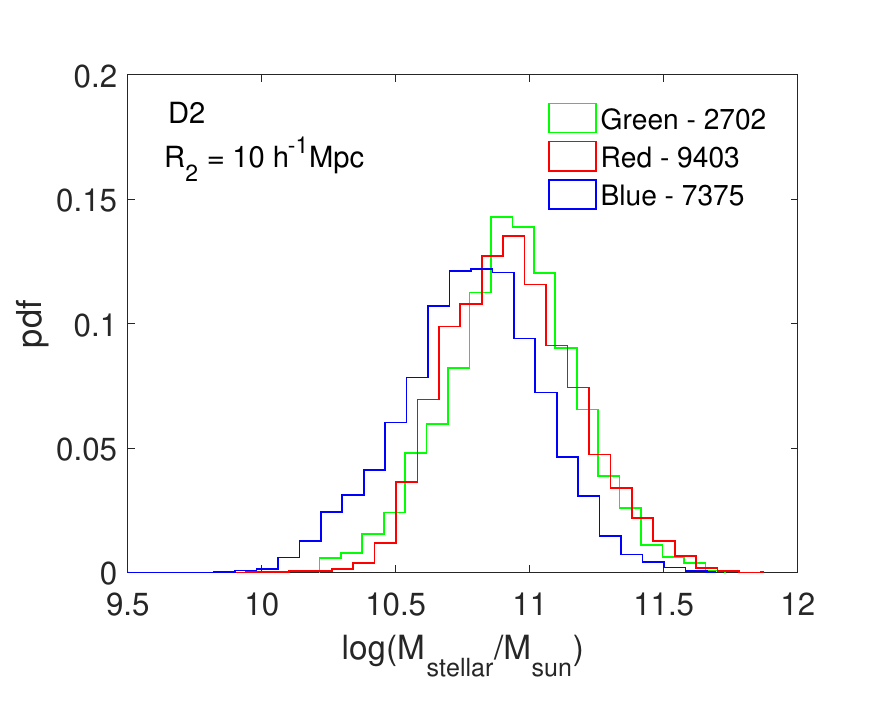}\hspace{0.1cm}
\vspace{-0.48cm}
\includegraphics[width=7cm]{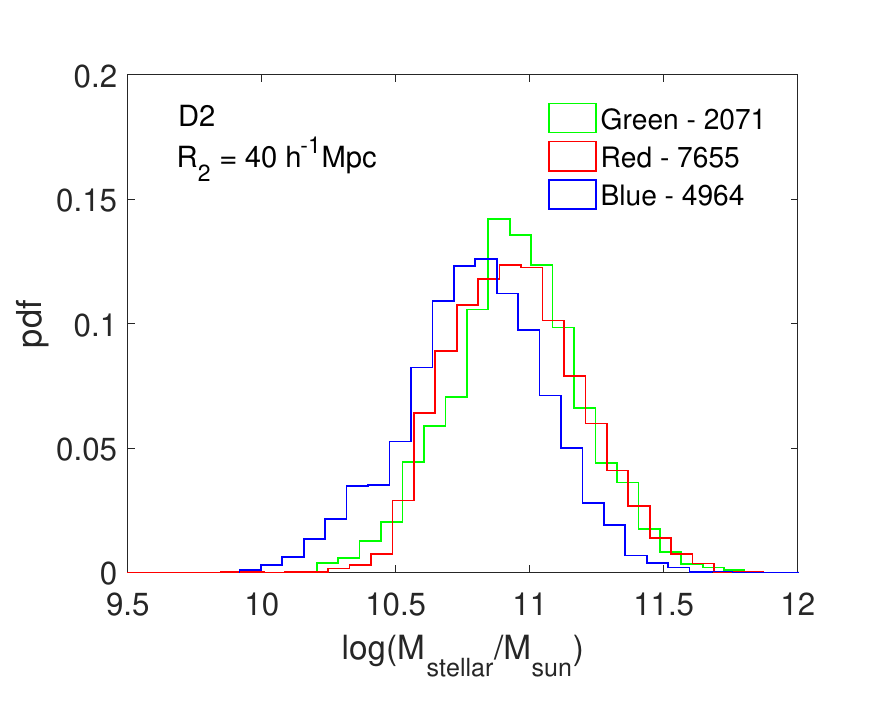}
\includegraphics[width=7cm]{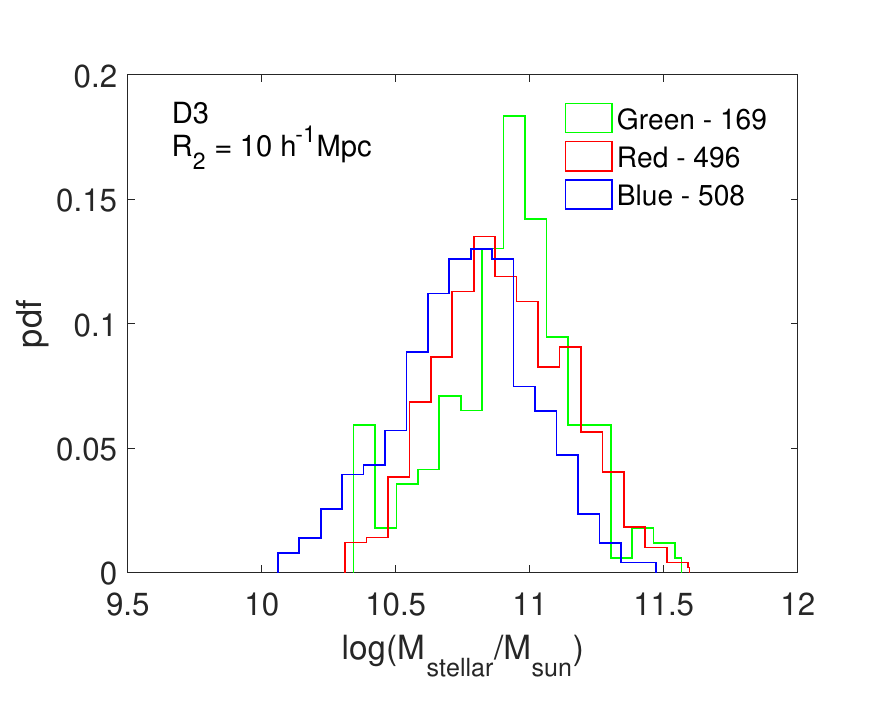}\hspace{0.1cm}
\includegraphics[width=7cm]{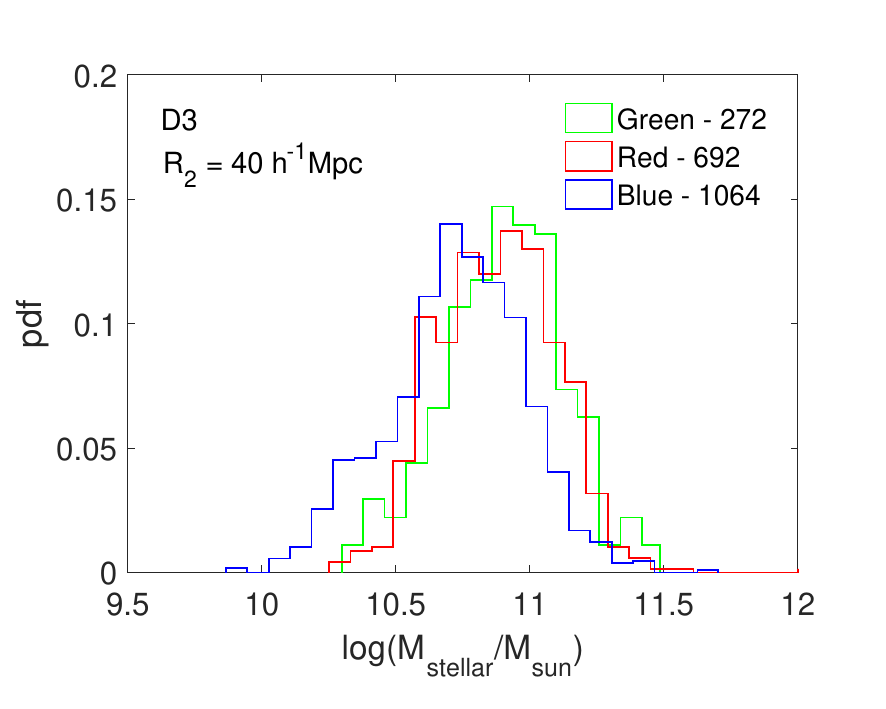}
\caption{The top left, middle left and bottom left panels show the
  stellar mass distributions of red, blue and green galaxies in,
  respectively $D1$, $D2$ and $D3$-type environments for $R_2=10
  \hmpc$. The same for $R_2=40 \hmpc$ are shown in the top right,
  middle right and bottom right panels of the figure. The number of
  red, blue and green galaxies available in each case are mentioned in
  the respective panels.}
\label{fig:mass1}
\end{figure*}

\begin{figure*}[htbp!]
\centering
\includegraphics[width=7cm]{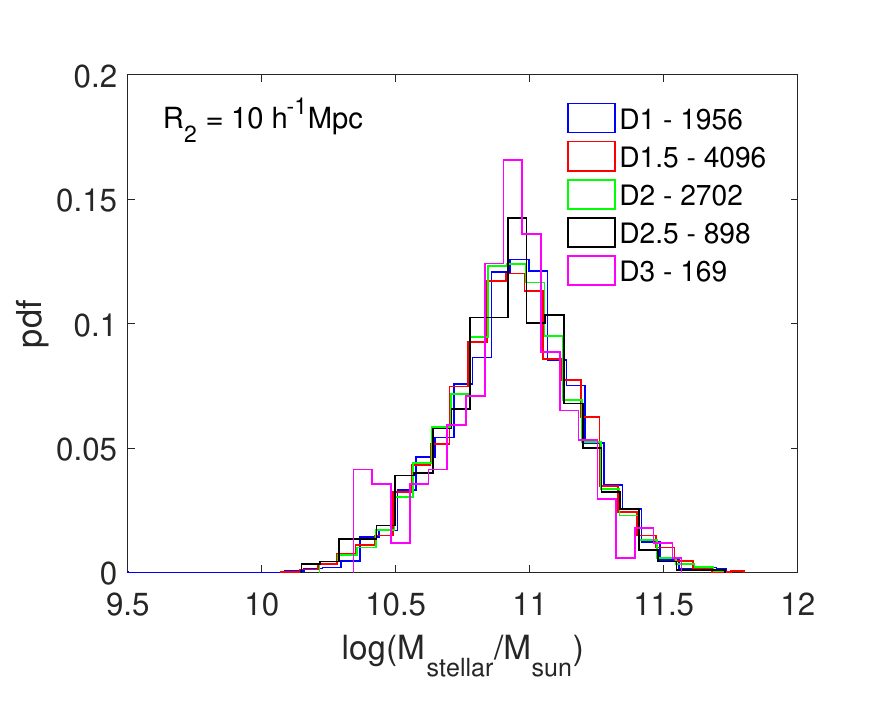}\hspace{0.1cm}
\vspace{-0.48cm}
\includegraphics[width=7cm]{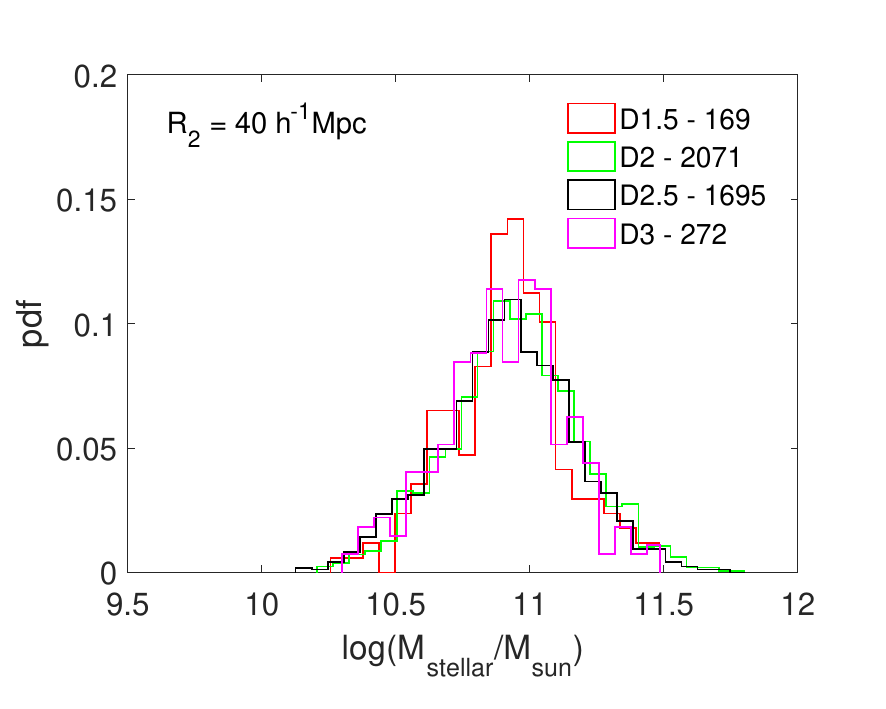}
\includegraphics[width=7cm]{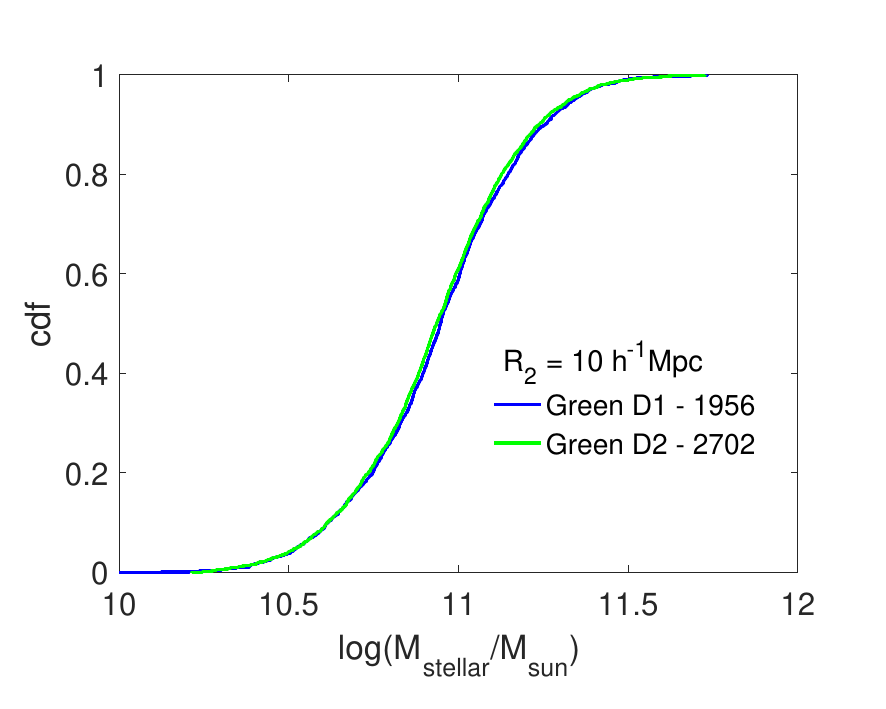}\hspace{0.1cm}
\vspace{-0.48cm}
\includegraphics[width=7cm]{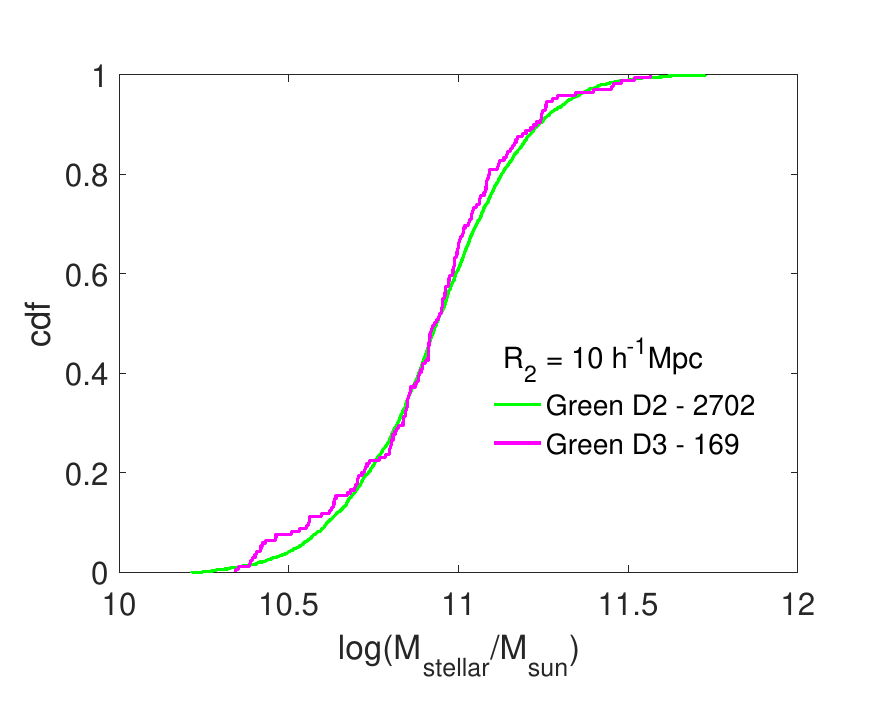}
\includegraphics[width=7cm]{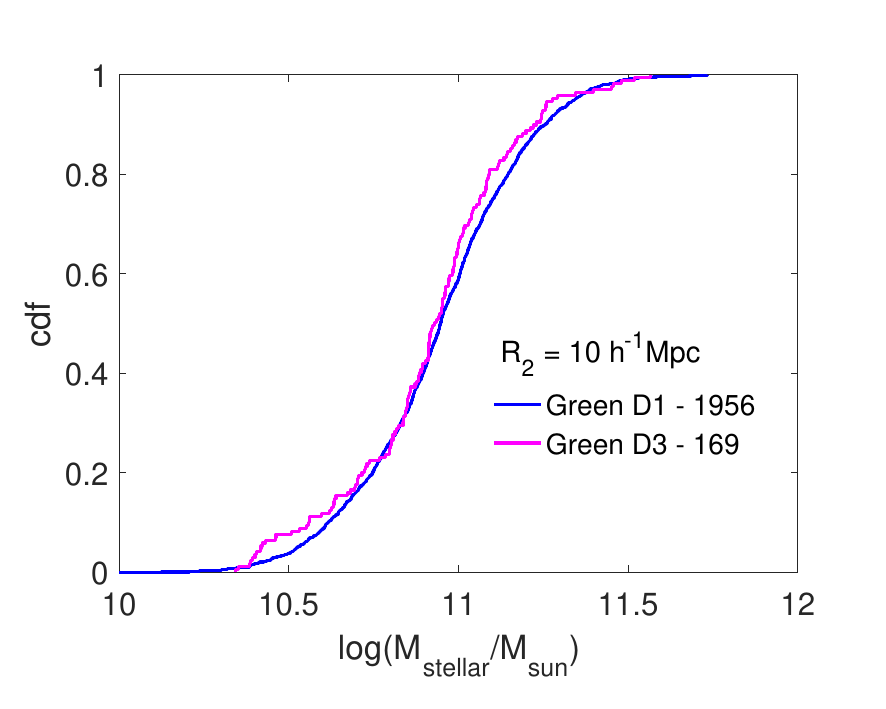}\hspace{0.1cm}
\includegraphics[width=7cm]{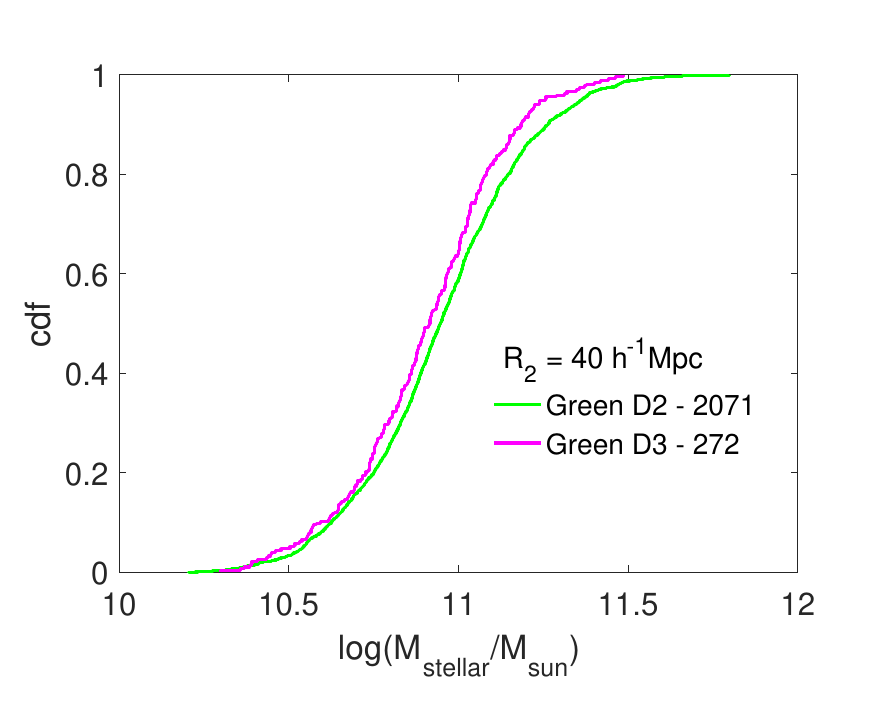}
\caption{The top left and top right panels compare the stellar mass
  distribution of green galaxies in different environments for $R_2=10
  \hmpc$ and $R_2=40 \hmpc$ respectively. The number of green galaxies
  available at different environments are mentioned in the top two
  panels. The middle and bottom panels compare the CDF of the stellar
  mass distribution for the green galaxies. The type of environments
  between which the comparison is made and the associated length
  scales are mentioned in each panel.}
\label{fig:mass2}
\end{figure*}

\subsection{Distributions of stellar mass in different environments}

In the top left, middle left and bottom left panels of
\autoref{fig:mass1}, we show the distributions of the stellar mass for
the red, blue and green galaxies in $D1$-type, $D2$-type and $D3$-type
environments with $R_2=10 \hmpc$. The results for $R_2=40 \hmpc$ in
$D1.5$-type, $D2$-type and $D3$-type environments are respectively
shown in the top right, middle right and bottom right panels of
\autoref{fig:mass1}. We see that at each environment, the
distributions of stellar mass of the red and green galaxies are quite
similar and are noticeably different than that for the blue
galaxies. The stellar mass distribution of the blue galaxies extends
to lower masses compared to the red and green galaxies in each type
of environment.

We also compare the stellar mass distributions of green galaxies in
different environments in the top two panels of
\autoref{fig:mass2}. The left and right panels in this figure
correspond to $R_2=10 \hmpc$ and $R_2=40 \hmpc$ respectively. We test
the statistical significance of the differences between the
distributions using a two-sample Kolmogorov-Smirnov test. The middle
and bottom panels of \autoref{fig:mass2} compare the CDF of stellar
mass distribution in different environments. The $D_{KS}$ values from
the test listed in \autoref{tab:dks}, suggests that at $R_2=10 \hmpc$,
the null hypothesis can not be rejected at high confidence level. So
there are no statistically significant difference between the stellar
mass distributions of green galaxies in different environments.However
we note that the stellar mass distribution of green galaxies in $D2$
and $D3$ type environments differ at $90\%$ confidence level at
$R_2=40 \hmpc$.

\begin{figure*}[htbp!]
\centering
\includegraphics[width=7cm]{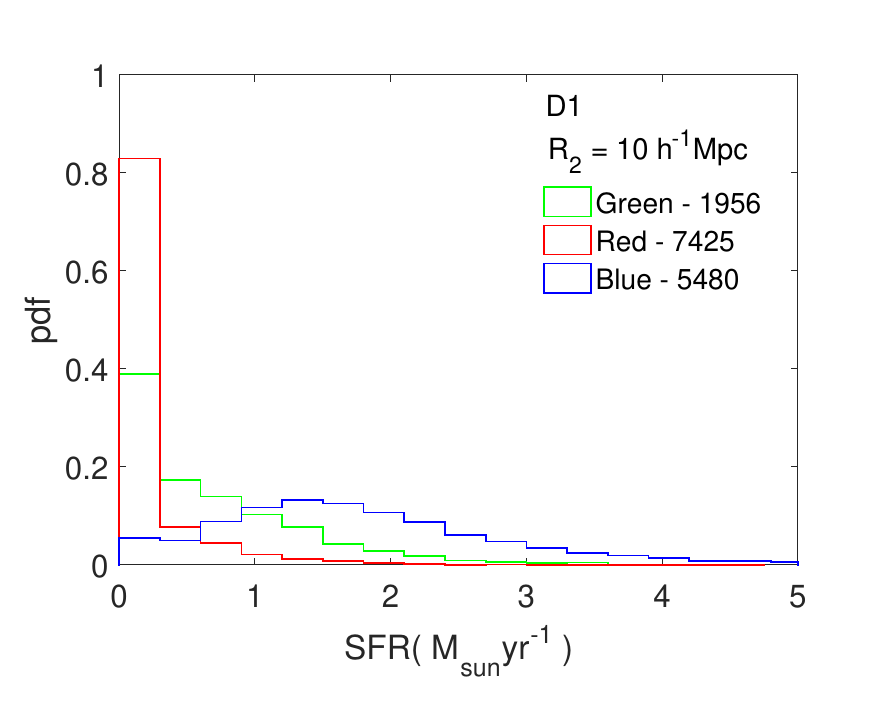}\hspace{0.1cm}
\vspace{-0.49cm}
\includegraphics[width=7cm]{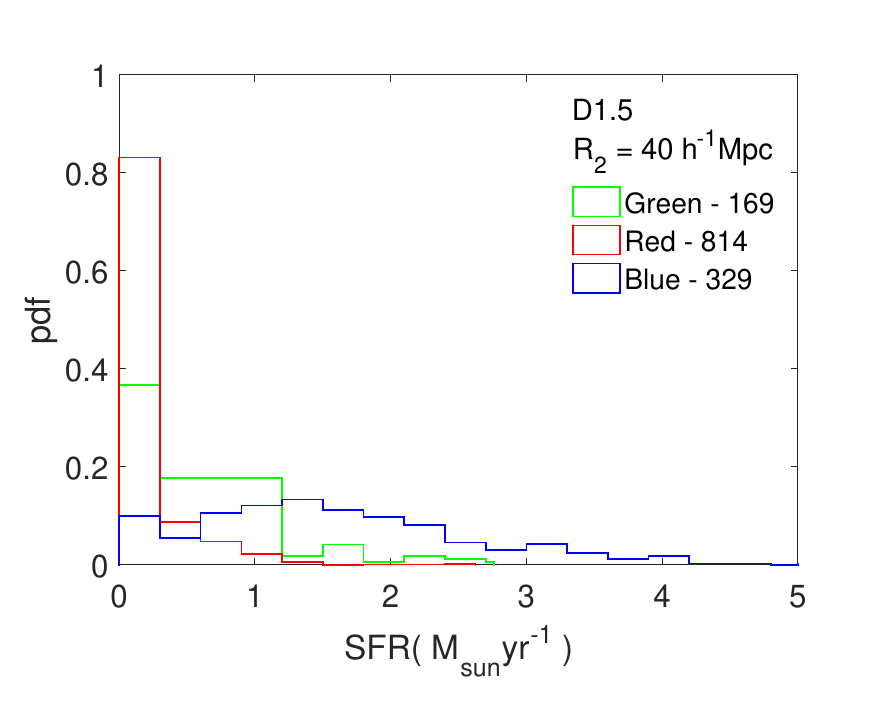}
\includegraphics[width=7cm]{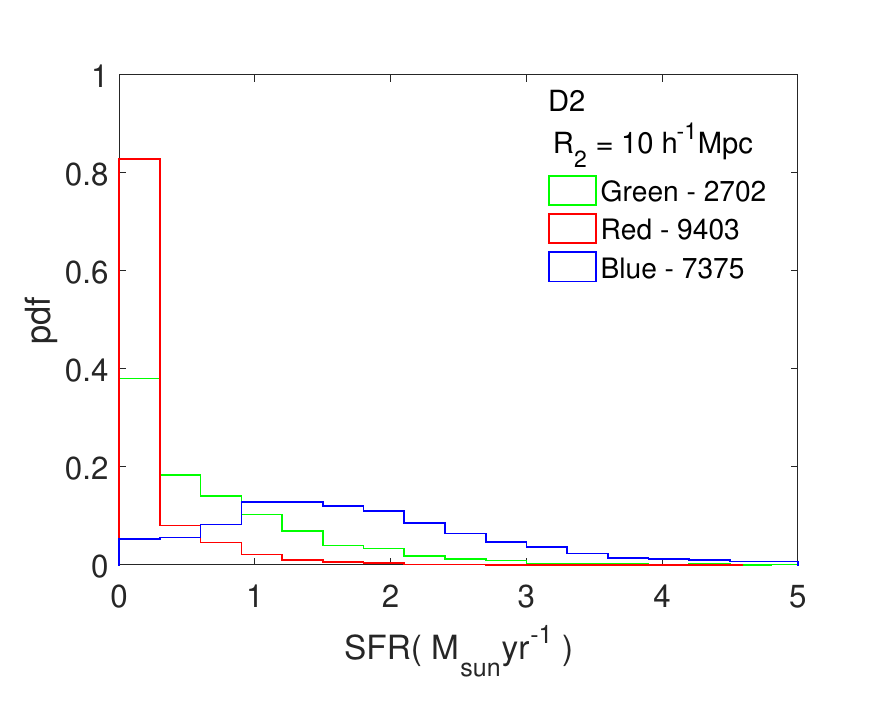}\hspace{0.1cm}
\vspace{-0.49cm}
\includegraphics[width=7cm]{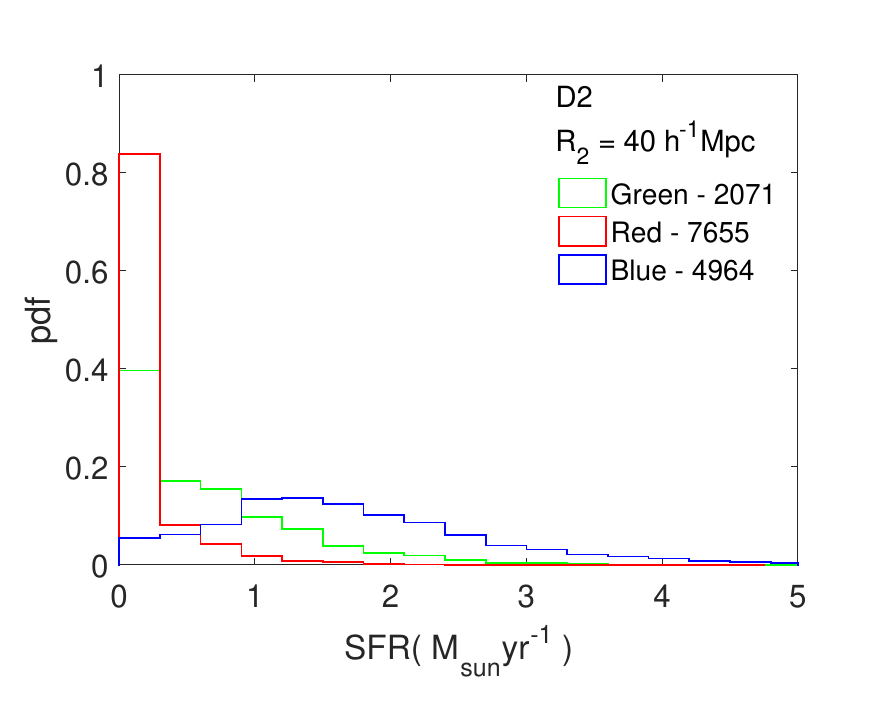}
\includegraphics[width=7cm]{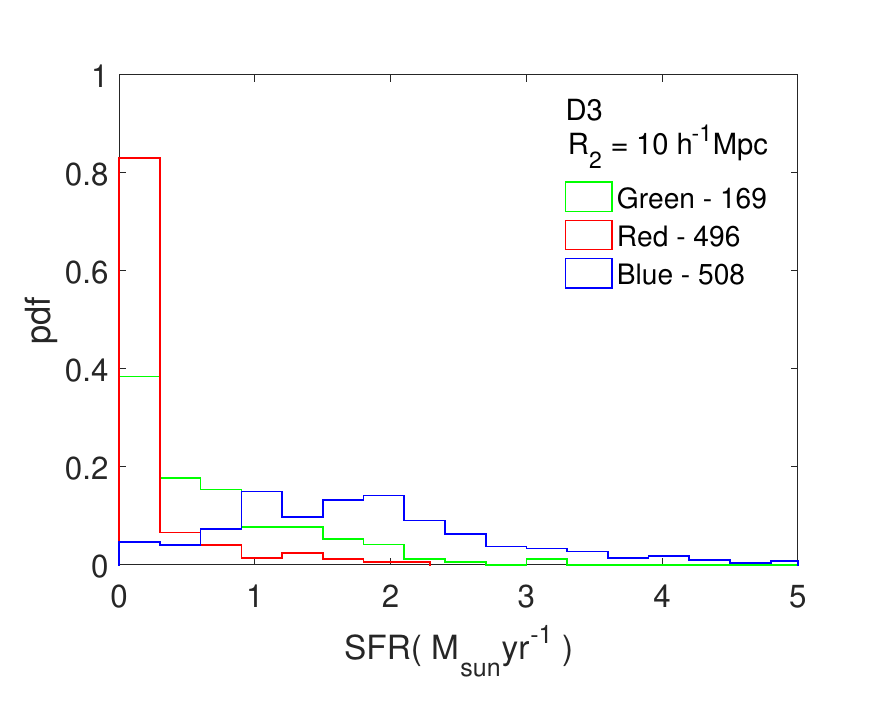}\hspace{0.1cm}
\includegraphics[width=7cm]{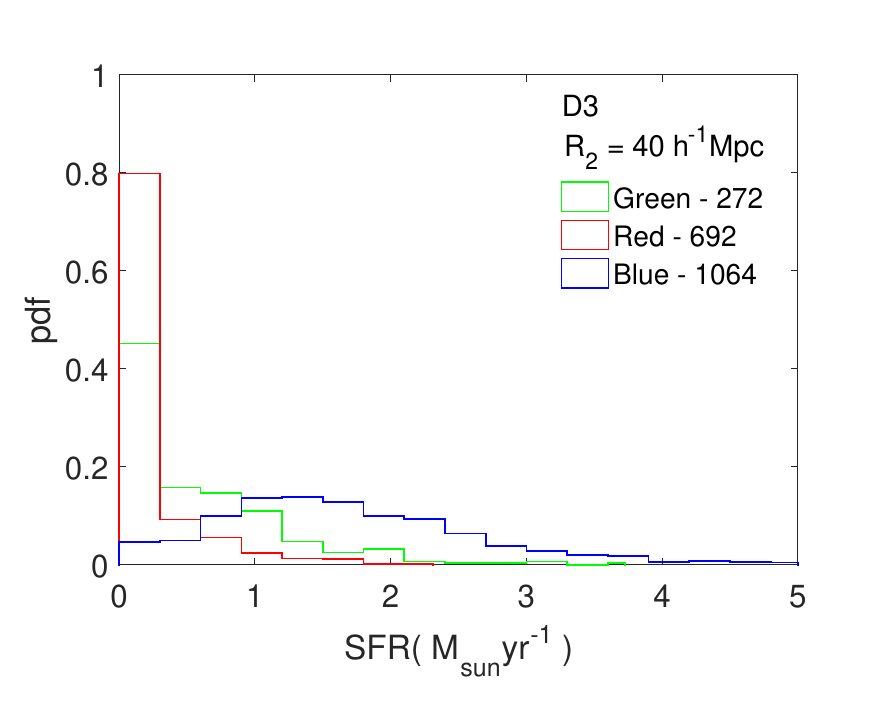}
\caption{Same as \autoref{fig:mass1} but for star formation rate.}
\label{fig:sfr1}
\end{figure*}

\begin{figure*}[htbp!]
\centering
\includegraphics[width=7cm]{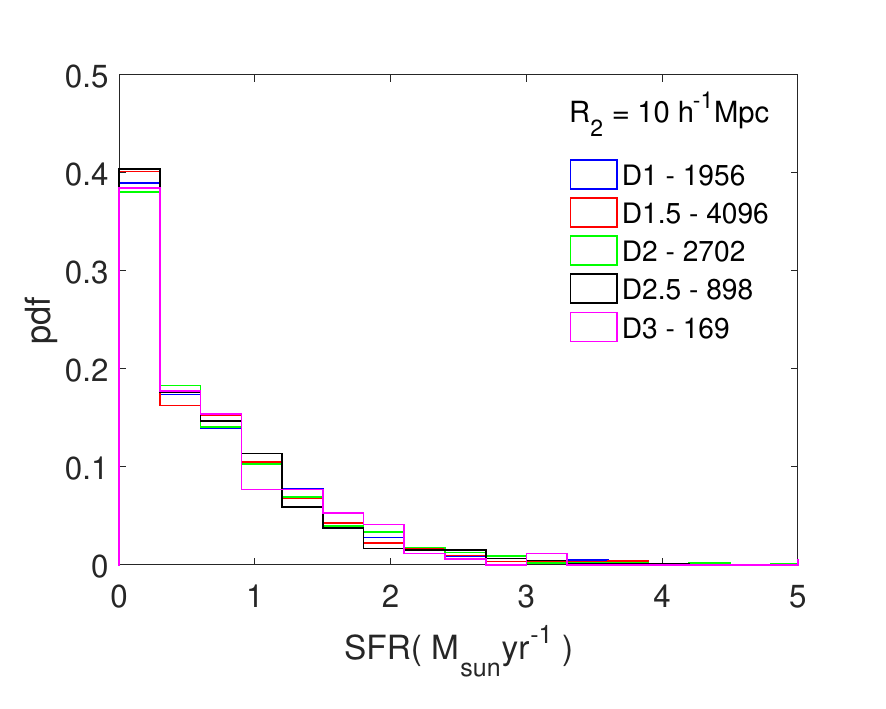}\hspace{0.1cm}
\vspace{-0.49cm}
\includegraphics[width=7cm]{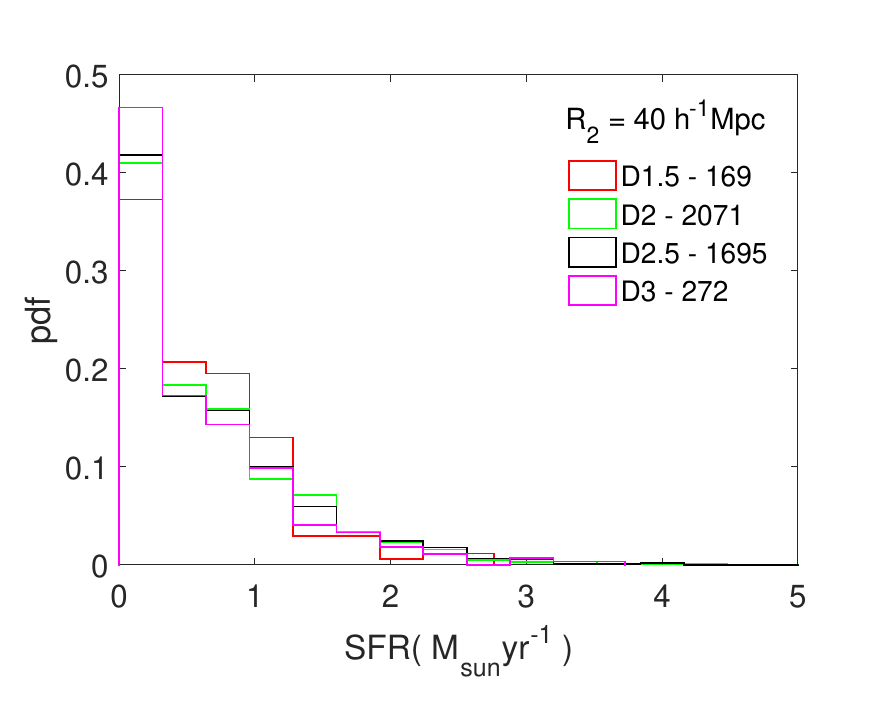}
\includegraphics[width=7cm]{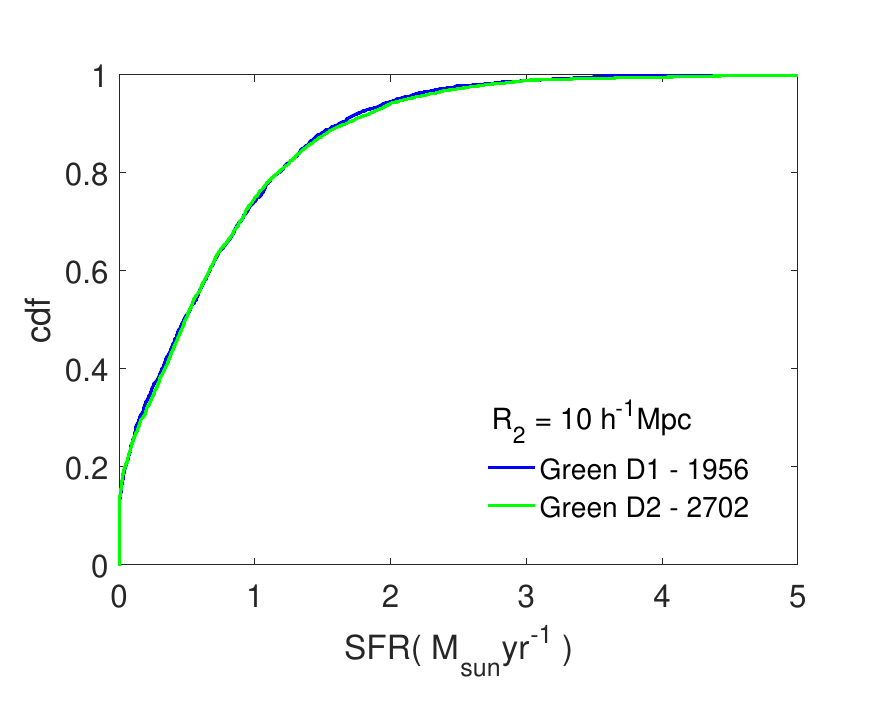}\hspace{0.1cm}
\vspace{-0.49cm}
\includegraphics[width=7cm]{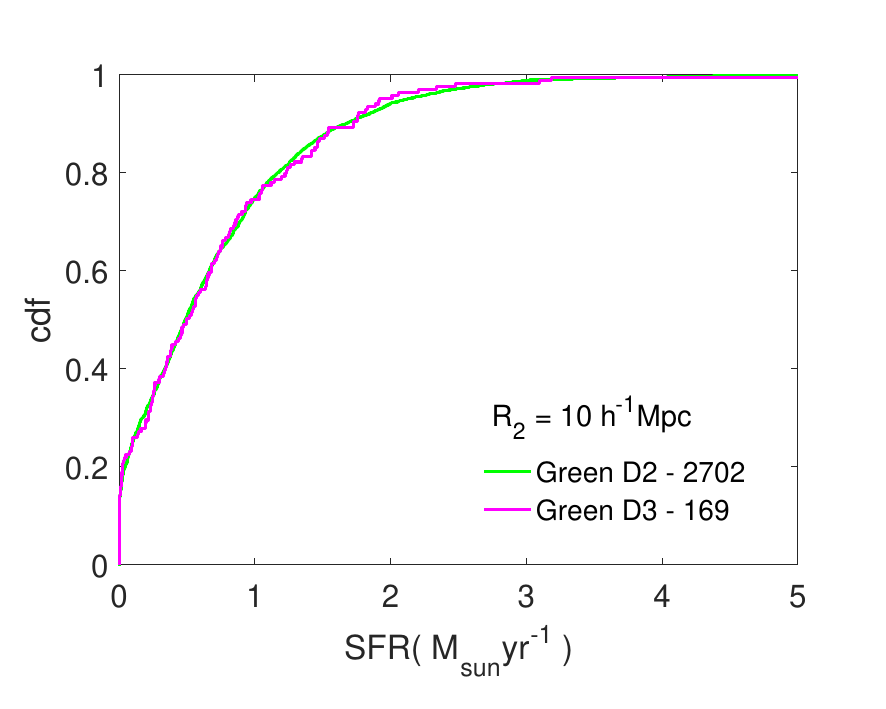}
\includegraphics[width=7cm]{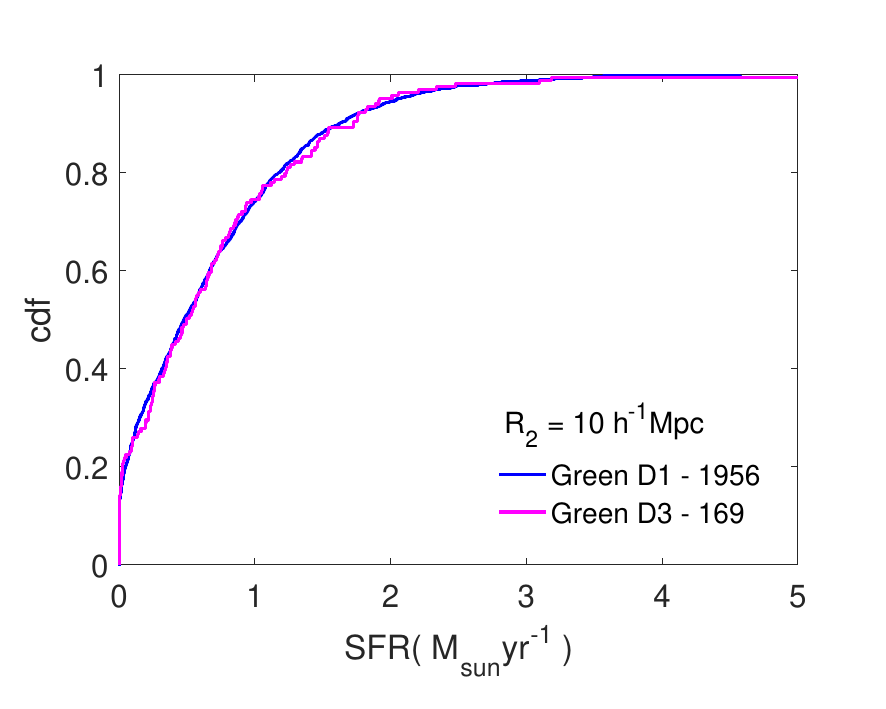}\hspace{0.1cm}
\includegraphics[width=7cm]{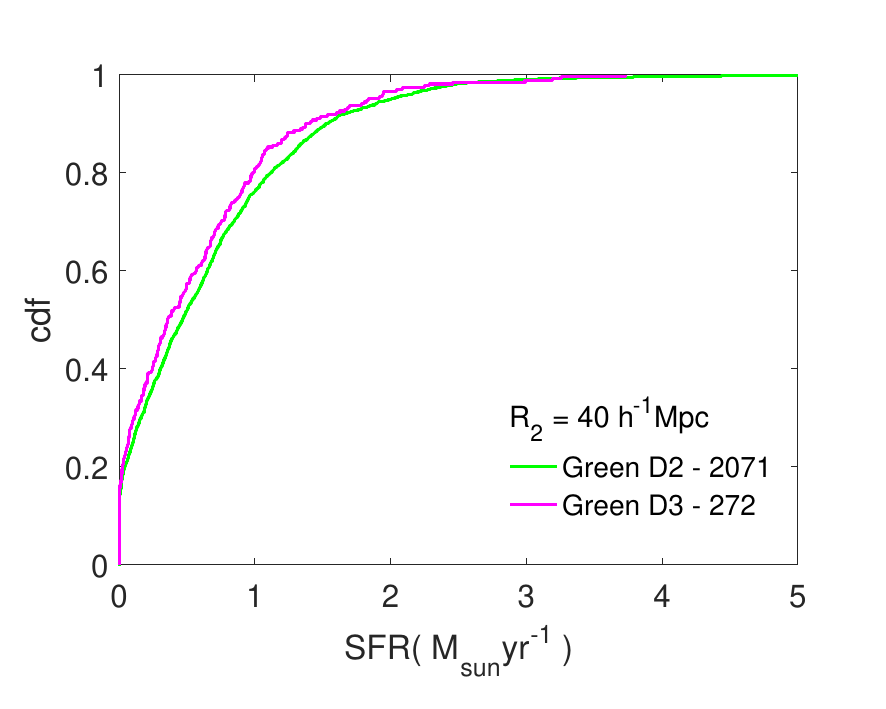}
\caption{Same as \autoref{fig:mass2} but for star formation rate.}
\label{fig:sfr2}
\end{figure*}

\subsection{Distributions of star formation rate in different environments}

We show the distributions of star formation rate (SFR) in red, blue
and green galaxies in different types of environment in
\autoref{fig:sfr1}. The left and right panels of \autoref{fig:sfr1}
show the results for $R_2=10 \hmpc$ and $R_2=40 \hmpc$
respectively. The SFR distribution of red galaxies at each environment
peaks at the lowest SFR bin. The peak of the SFR distribution of green
galaxies occur at the lowest SFR bin too. However, we see that $\sim
80\%$ of the red galaxies and $\sim 40\%$ green galaxies reside in the
lowest SFR bin. So there is a large difference between the amplitudes
of the peaks associated with the two distributions. The distribution
of green galaxies dominate the distribution of red galaxies in all the
other SFR bins. The SFR distribution of green galaxies extends to
higher SFR values as compared to the distribution of red galaxies. The
distribution of blue galaxies peaks at higher SFR and extends further
compared to that for the red and green galaxies at each environment.

We compare the SFR distributions of the green galaxies in different
types of environments for $R_2=10 \hmpc$ and $R_2=40 \hmpc$ in the top
two panels of \autoref{fig:sfr2}. We test the statistical significance
of the differences between the distribution using a Kolmogorov-Smirnov
test. The CDFs in different environments are compared in the middle
and bottom panels of \autoref{fig:sfr2} and the corresponding $D_{KS}$
values are listed in \autoref{tab:dks}. The results show that the null
hypothesis can not be rejected even at $80\%$ confidence level which
implies that the SFR in green galaxies are independent of their
environment.

\begin{figure*}[htbp!]
\centering
\includegraphics[width=7cm]{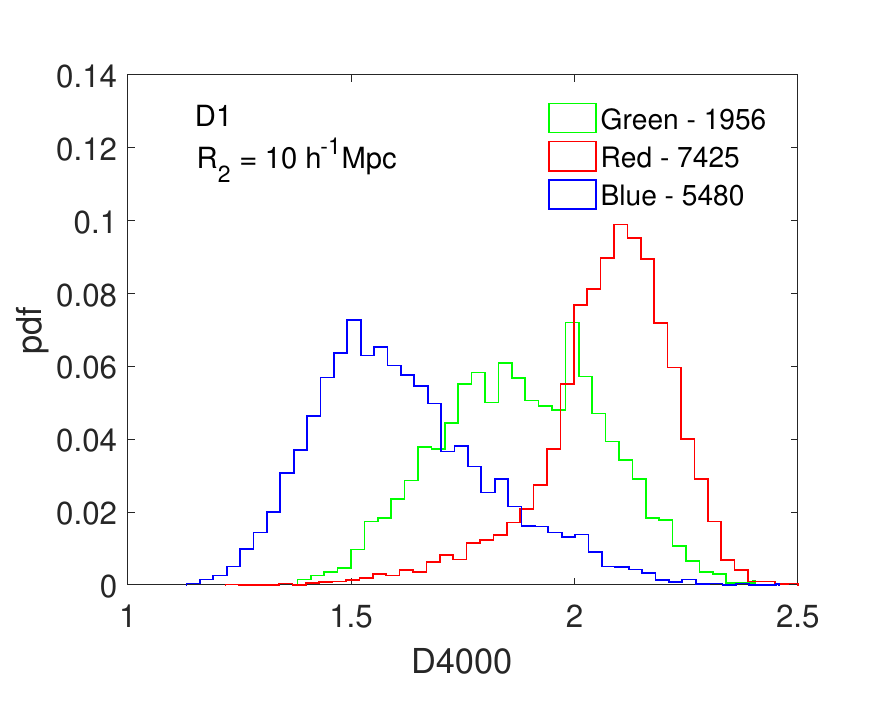}\hspace{0.1cm}
\vspace{-0.49cm}
\includegraphics[width=7cm]{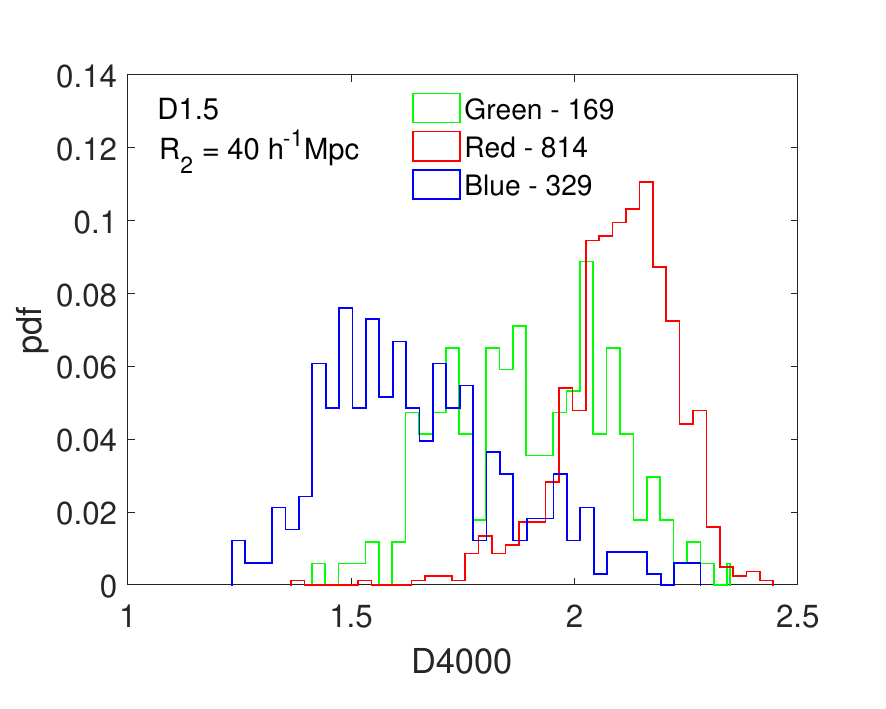}
\includegraphics[width=7cm]{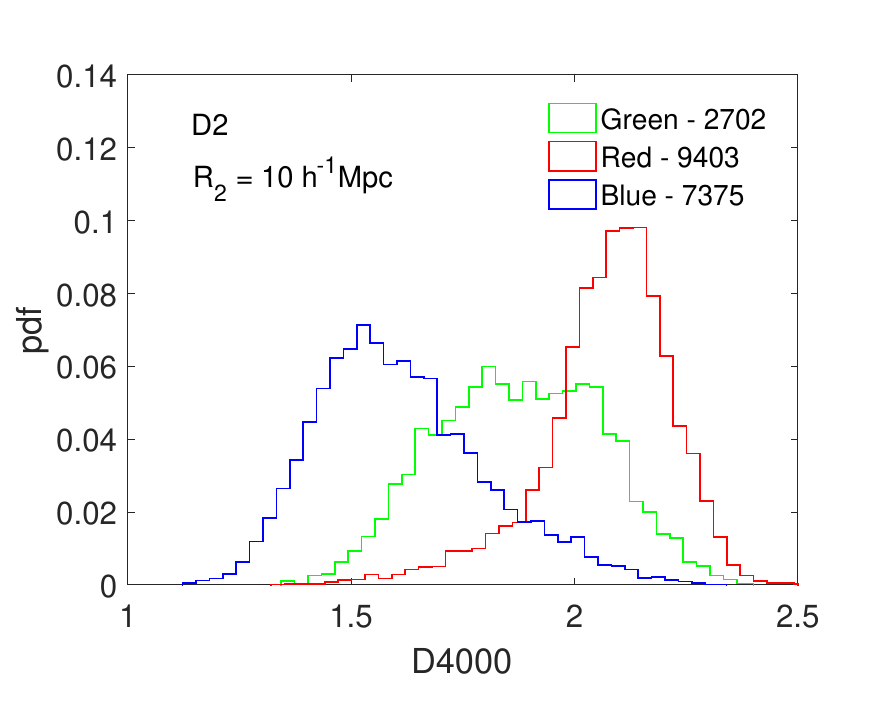}\hspace{0.1cm}
\vspace{-0.49cm}
\includegraphics[width=7cm]{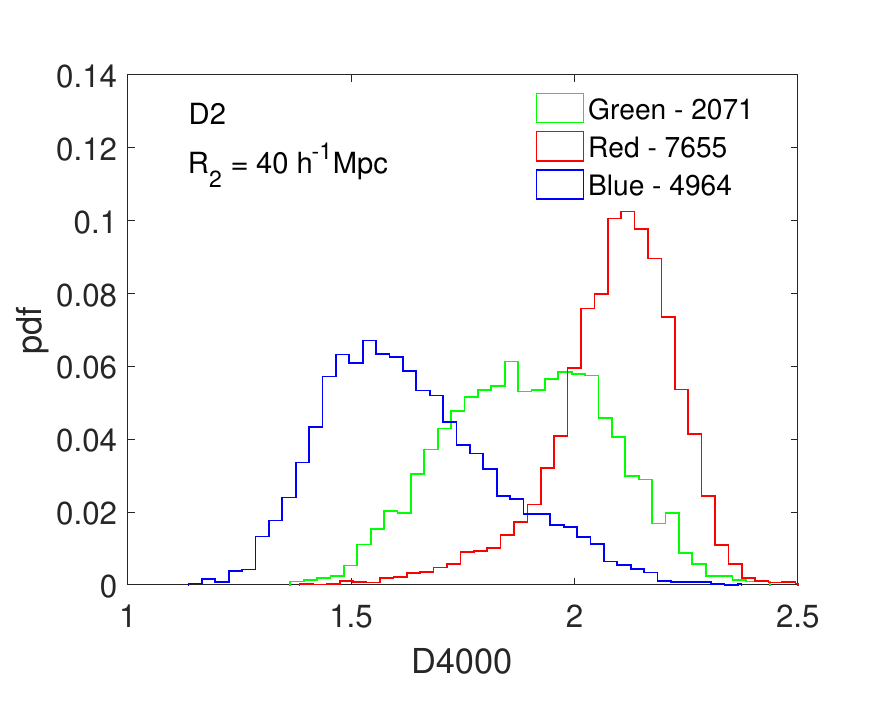}
\includegraphics[width=7cm]{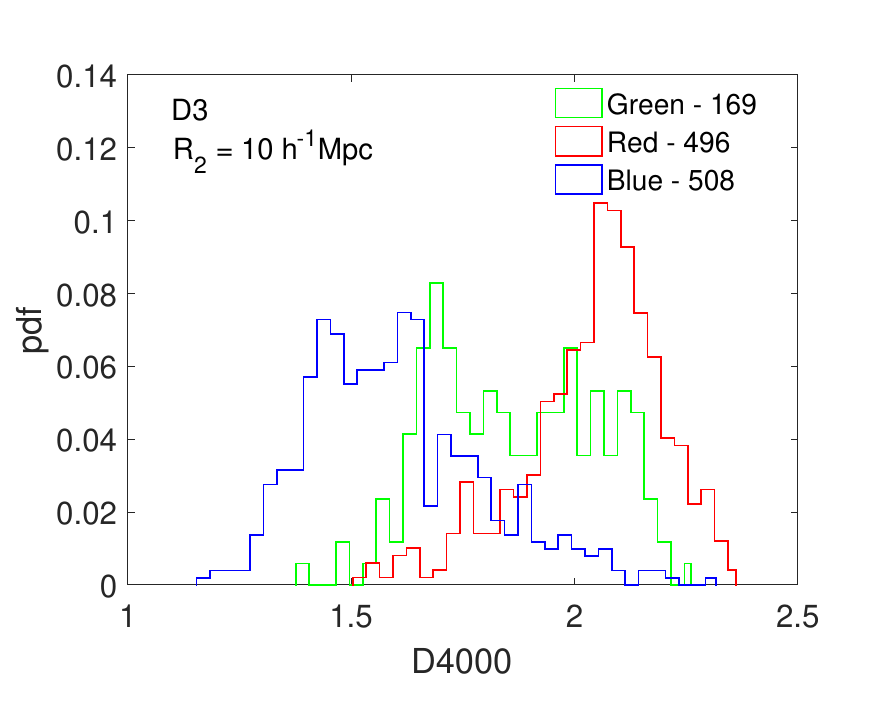}\hspace{0.1cm}
\includegraphics[width=7cm]{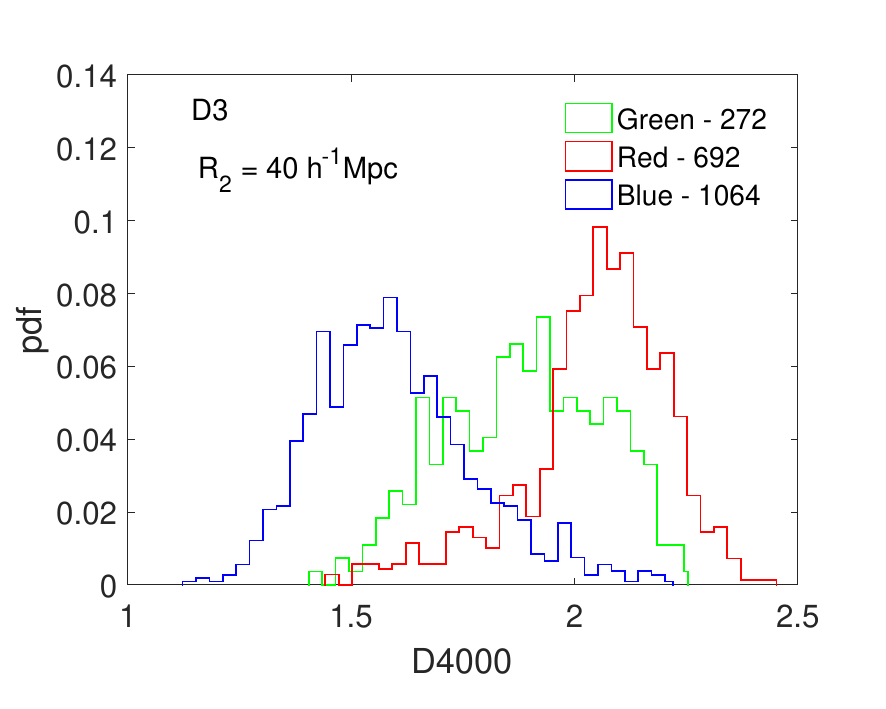}
\caption{Same as \autoref{fig:mass1} but for stellar age.}
\label{fig:age1}
\end{figure*}

\begin{figure*}[htbp!]
\centering
\includegraphics[width=7cm]{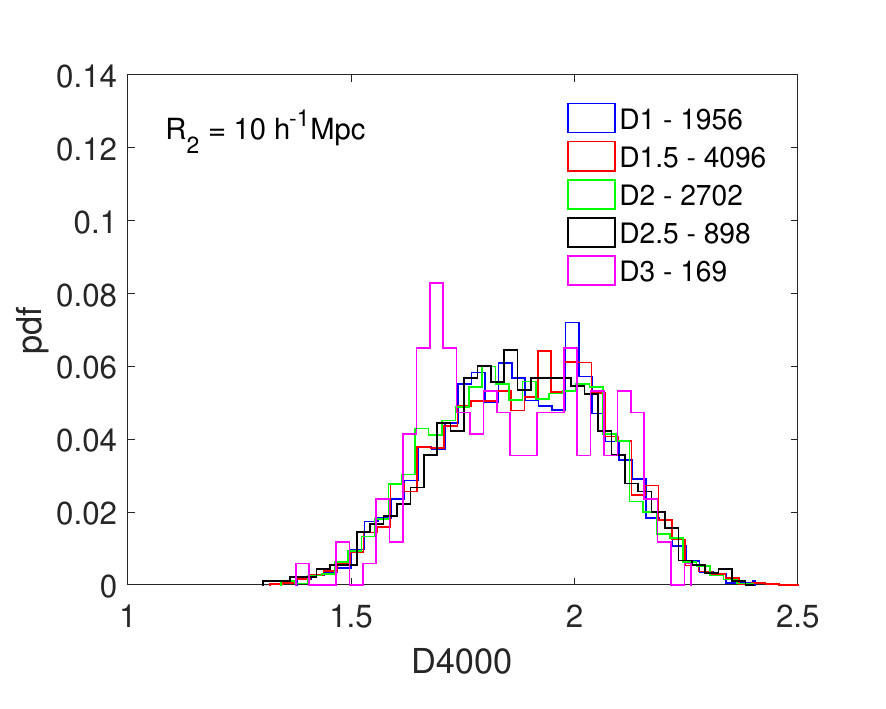}\hspace{0.1cm}
\vspace{-0.49cm}
\includegraphics[width=7cm]{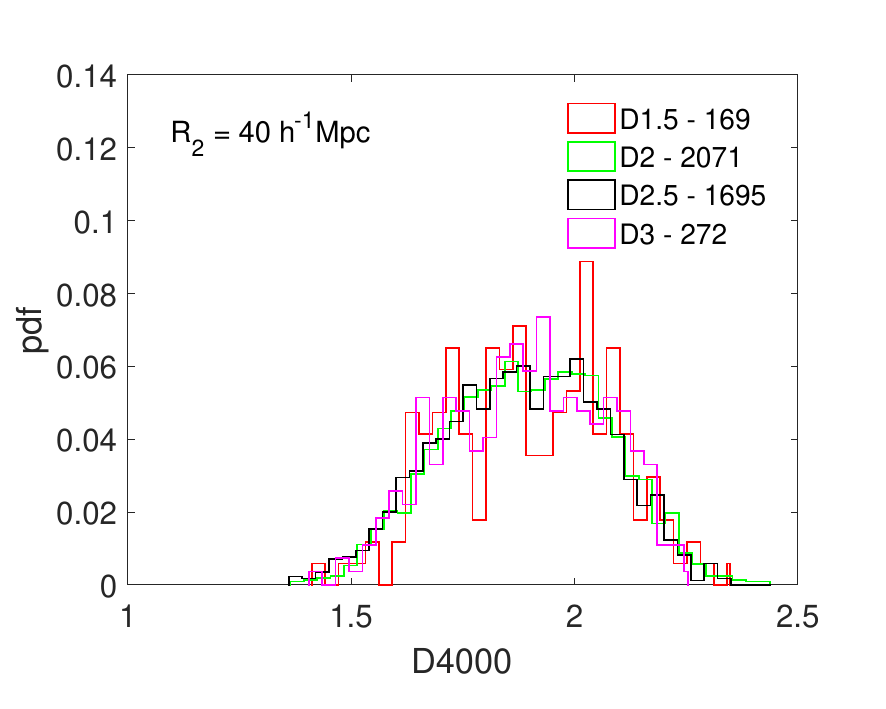}
\includegraphics[width=7cm]{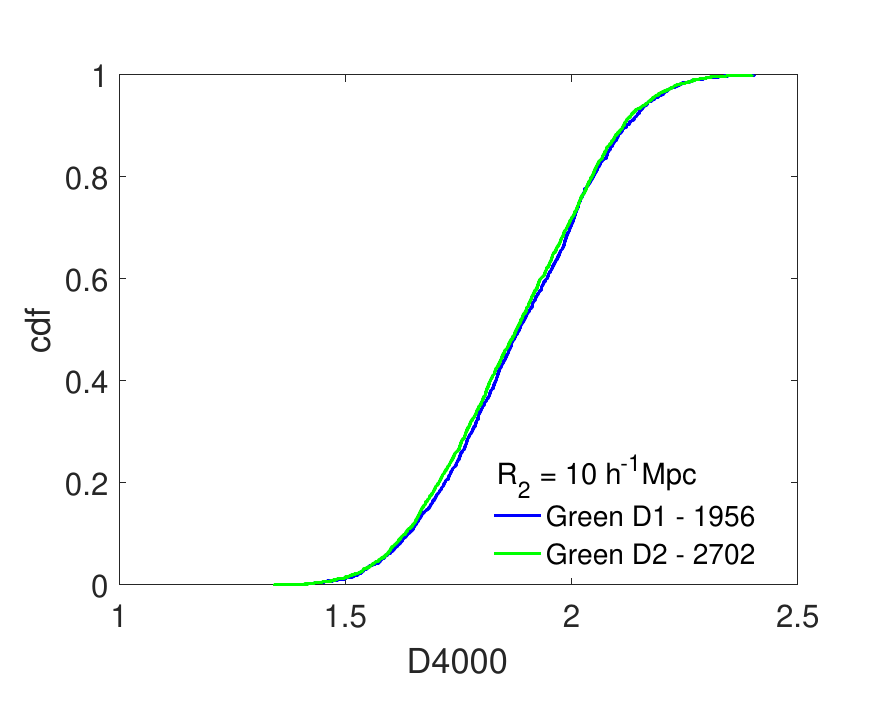}\hspace{0.1cm}
\vspace{-0.49cm}
\includegraphics[width=7cm]{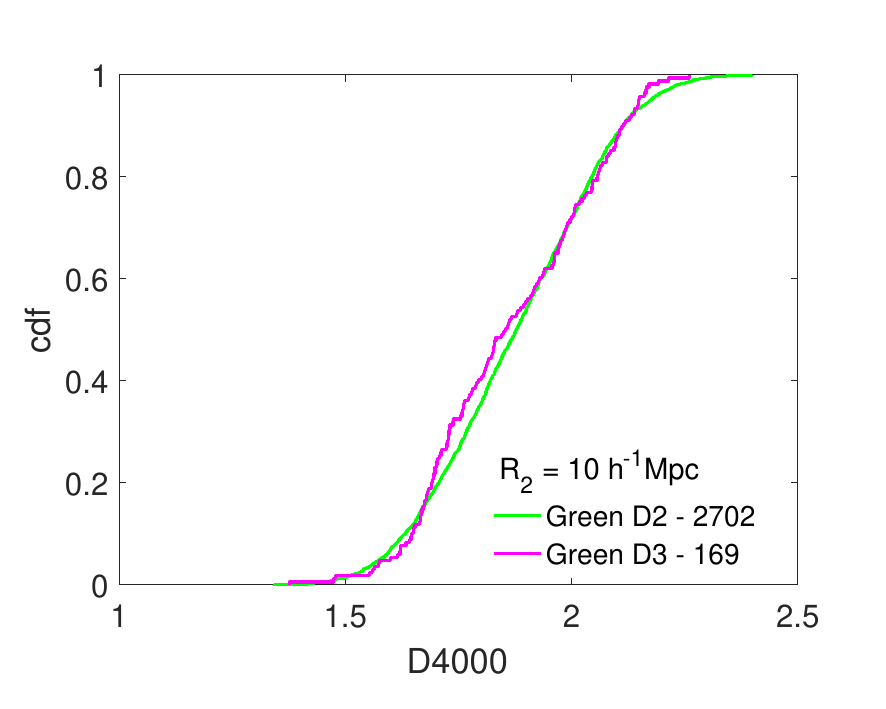}
\includegraphics[width=7cm]{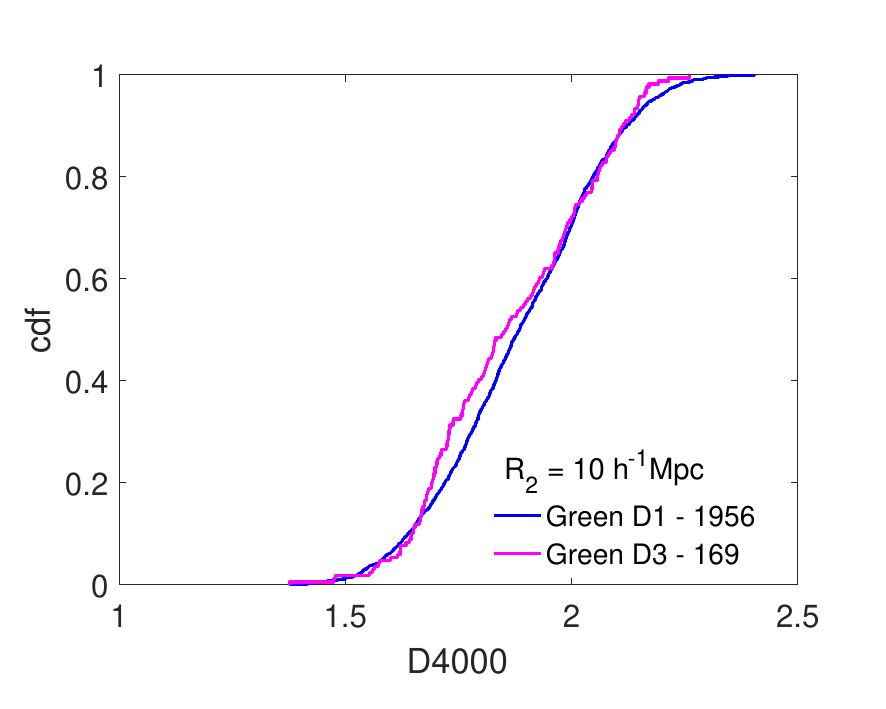}\hspace{0.1cm}
\includegraphics[width=7cm]{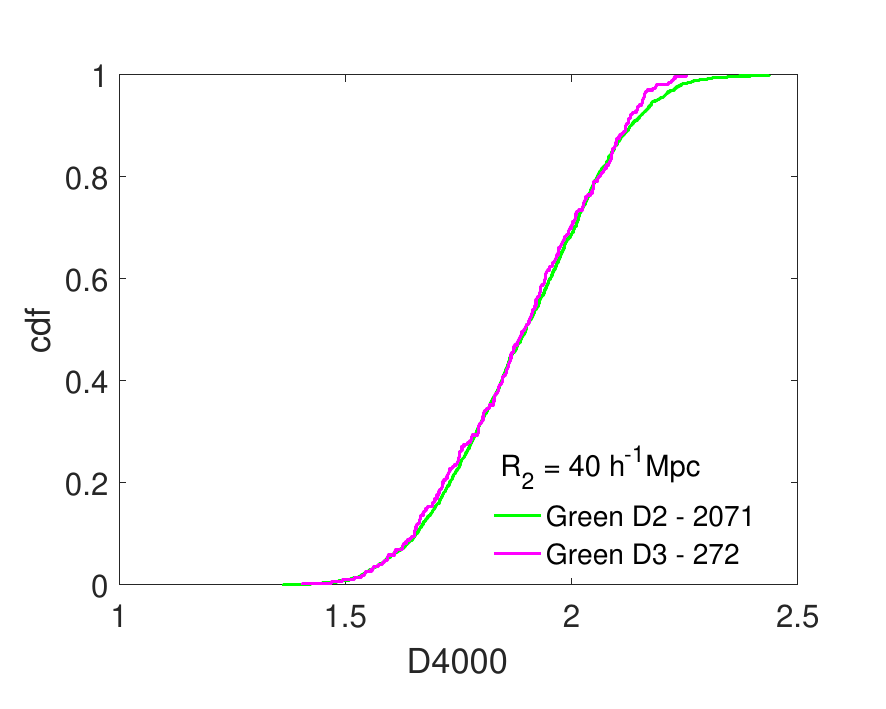}
\caption{Same as \autoref{fig:mass2} but for stellar age.}
\label{fig:age2}
\end{figure*}

\subsection{Distributions of stellar age in different environments}

We show the distributions of stellar age as characterized by $D4000$
for the red, blue and green galaxies in different environments in
\autoref{fig:age1}. The top, middle and bottom left panels of
\autoref{fig:age1} show the results for $R_2=10 \hmpc$. The results
for $R_2=40 \hmpc$ are shown in the top, middle and bottom right
panels of \autoref{fig:age1}. At each panel, the distributions of the
red and blue galaxies show two distinct peaks at $D4000=2.1$ and
$D4000=1.5$ respectively. The wide separation between the two peaks
suggests that the red galaxies are dominated by older stellar
population whereas the blue galaxies are primarily represented by
younger stellar population.  The distribution of the green galaxies
show a broader peak at the intermediate region between the two peaks
corresponding to the red and blue galaxies. This suggests that the age
of the stellar population hosted in green galaxies are relatively
younger than red galaxies and older than blue galaxies. The
distribution for the green galaxies also show a significant overlap
with the distributions of red and blue galaxies.

A comparison of the stellar age distributions for the green galaxies
in different environments are shown in the top two panels of
\autoref{fig:age2}. The statistical significance of the differences
between the distributions are measured using a Kolmogorov-Smirnov
test. We compare the CDFs in the middle and bottom panels of
\autoref{fig:age2} and list the corresponding $D_{KS}$ values in
\autoref{tab:dks}. The test shows that the null hypothesis can not be
rejected at high confidence level which indicates that the
stellar age of green galaxies are independent of their environment. 




\begin{table}
	\centering
\begin{tabular}{|c|c|ccc|c|}
\hline
& &\multicolumn{3}{c|}{$R_2 = 10 \hmpc$}& \multicolumn{1}{c|}{$R_2 = 40 \hmpc$}\\
 
& & D1-D2 & D2-D3 & D1-D3 & D2-D3\\ 
\hline
& $log(M_{stellar}/M_{sun})$ & 0.0284 & 0.0568 & 0.0733 & 0.0892\\
$D_{KS}$ & SFR & 0.0220 & 0.0328 & 0.0532 & 0.0665\\
 & $D4000$ & 0.0271 & 0.0760 & 0.0969 & 0.0369\\
\hline
                 & 99\% & 0.0483 & 0.1291 & 0.1305 & 0.1050\\
                 & 90\% & 0.0363 & 0.0970 & 0.0981 & 0.0789\\
$D_{KS}(\alpha)$ & 80\% &  0.0319 & 0.0851 & 0.0860 & 0.0692\\
                 & 70\% & 0.0289 & 0.0722 & 0.0781 & 0.0628\\
                 & 60\% & 0.0266 & 0.0711 & 0.0719 & 0.0579\\
                 & 50\% & 0.0247 & 0.0665 & 0.0668 & 0.0537\\

\hline
\end{tabular}

\caption{This table shows the Kolmogorov-Smirnov statistic $D_{KS}$
  for comparisons of $\log(M_{stellar}/M_{sun})$, $SFR$ and $D4000$
  distributions of green galaxies in various environments for $R_2=10
  \hmpc$ and $R_2=40 \hmpc$.  The table also lists the critical values
  of $D_{KS}(\alpha)$ above which null hypothesis can be rejected at
  different confidence levels.}
\label{tab:dks}
\end{table}

\section{Conclusions}
In the present work, we analyze a volume limited sample from the SDSS
to understand the internal and external influences on the quenching of
star formation in the green valley galaxies. We employ a fuzzy
set-theoretic method to classify the galaxies according to their
colour and characterize their environment using the local dimension.
We study the fraction of green galaxies in different environments as a
function of their stellar mass. We also study the morphology of green
galaxies in different environments. The AGN driven winds can quench
star formation in galaxies. They can also play an important role in
mass driven quenching. So, we study the fraction of AGNs in green
galaxies in different environments. The presence of a dominant bulge
or bar can also suppress star formation in galaxies. We measure the
fraction of green galaxies with dominant bulge and bar in various
environments. Galaxy interaction such as strangulation and harassment
are also known to quench star formation in galaxies. We study the
fraction of green galaxies with disturbed, irregular or merger
features in different environments. We finally compare the
distributions of stellar mass, SFR and stellar age of red, blue and
green galaxies in different environments of the cosmic web. We test if
the distributions of stellar mass, SFR and stellar age of green
galaxies in various environments have any statistically significant
differences. Our main results can be summarized as follows:

(i) The fraction of green galaxies is independent of geometric
environments of the cosmic web. Around $10\%-20\%$ of the galaxies in
each environment turn out to be green. The fraction of the green
galaxies increases with the stellar mass but remains below $20\%$ at
each environment. The green fraction is also independent of the length
scales associated with the environments.

(ii) At each environment, $\sim 10\%$ of green galaxies show AGN
activity. The fraction of AGN in green galaxies is also independent of
the environment and the associated length scales.

(iii) At each environment, $\sim 95\%$ of the visually classified
green galaxies are spirals and only $\sim 5\%$ are ellipticals. The
morphology of the green galaxies is also independent of the
environment and the associated length scales.

(iv) Most of the visually classified green galaxies have a bulge but
only $\sim 1\%$ of them at each environment, have a dominant
bulge. Nearly $6\%-8\%$ of classified green galaxies at each
environment host a bar. The bulge fraction and bar fraction in green
galaxies show a mild environmental dependence with no clear trends.

(v) Only $\sim 4\%$, $\sim 3\%$ and $\sim 1\%$ visually classified green
galaxies at each environment respectively show disturbed, irregular
and merger features. The green fractions with these odd features show a
mild environmental dependence but no obvious trends are observed.

(vi) The stellar mass distributions of red and green galaxies are
quite similar in each environment. The distributions of SFR and
stellar age of red, blue and green galaxies are noticeably different
in each environment.

(vii) The distributions of stellar mass, SFR and stellar age of green
galaxies in various environments of the cosmic web do not differ in a
statistically significant way.

Based on the above findings, we conclude that the green valley
galaxies follow a mixture of evolutionary pathways. We do not find any
strong evidence in favour of environmental influences in quenching of
star formation in green galaxies. Only $\sim 8\%$ green galaxies in
each environment show signatures of interactions and merger. $5\%$ of
the green galaxies in each environment are ellipticals which may be
the outcome of gas-rich mergers. We also note that the distributions
of stellar mass, SFR and stellar age of green galaxies are independent
of their environments.

Majority of the green galaxies ($\sim 95\%$) at each environment are
disc dominated spirals. Contrary to this, most of the red galaxies are
known to be ellipticals with very little or no star formation. The
massive red ellipticals are believed to have acquired their mass
through a major merger or multiple minor mergers. The morphological
transformations in these galaxies were induced by mergers and their
higher masses may have played an important role in quenching star
formation in these galaxies. The similarity between the stellar mass
distributions of red and green galaxies and the fact that most green
galaxies are disc dominated spirals indicates that the vast majority
of green galaxies may have initiated their quenching due to their
larger masses. The disc dominated massive green galaxies may be an
outcome of secular evolution. Some galaxies at each environment may
grow only via smooth accretion and hence evolve very slowly. The star
formation in these galaxies is curtailed when their masses exceed the
critical mass required for the transition from cold to hot mode of
accretion. Once the hot mode is activated, the supply of cold gas from
the IGM to these galaxies are hindered. These galaxies slowly exhaust
their remaining cold gas over time and consequently show a decrease in
star formation. Presence of AGN activity can accelerate quenching of
star formation in these galaxies. But the fact that only $\sim 10\%$
green galaxies at each environment host AGNs indicates that they play
a minor role in suppressing the star formation in these
galaxies. Schawinski et al.\citep{schawinski09b} find that galaxies of
moderate-luminosity supermassive black holes in the local universe
have intermediate optical colors that imply the host galaxies are
transitioning from star formation to quiescence. Our results do not
contradict with these findings. These moderate-luminosity supermassive
black holes with intermediate optical colors may represent a certain
fraction of the green valley population for which the AGN activity
plays the central role in quenching. Further, Schawinski et
al.\citep{schawinski15} show that the time lag between the triggering
of AGN activity and the photoionization of the gas reservoir of the
galaxy is $\sim 10^5$ years. This is known to be much shorter than the
growth time of black holes ( $\sim 10^7-10^9$ years). It suggests that
the black holes may grow via many such on and off phases of AGN. If
such flickering of AGN activity are prevalent in the green valley
galaxies then it would be difficult to correctly evaluate the role of
AGNs in quenching the population. Our findings do not rule out such a
possibility.

Dynamical instabilities within a galaxy leading to the formation of
bar and bulge can also suppress star formation within a galaxy. The
presence of dominant bulge and bar in $\sim 10\%$ green galaxies at
each environment suggests that such internal influences may also play
a role of quenching agents in these galaxies. However, $\sim 70\%$ of
the green galaxies must curtail their star formation using physical
mechanism(s) other than interactions, mergers and those driven by
bulge, bar and AGN activity. We speculate that these green galaxies
are massive galaxies which have grown via secular evolution
\citep{kormendy04} and terminated their star formation via mass driven
quenching. However, a long term shutdown of star formation in these
galaxies requires the presence of additional heating sources like AGN
feedback, without which accretion and cooling of gas can not be
suppressed indefinitely. We observe that only a small fraction of
green galaxies show AGN activity and hence the quenching in green
galaxies is still a significant problem which requires other internal
physical processes to truncate the gas supply and maintain the hot
halo. We would also like to mention here that the present day green
valley galaxies do not represent the progenitors of the galaxies that
are currently in the red sequence. The progenitors of the red galaxies
were star-forming in the past and those galaxies were different than
the present day star-forming galaxies. So it is not clear whether the
galaxies in the red sequence have gone through a significant
morphological transformation or they were born more dispersion
dominated \citep{tacchella, bhattacharjee20}.

The incompleteness of our volume limited sample at low stellar masses
(\autoref{fig:hist2d}) is a caveat in the present analysis. We
construct the sample in order to have the local dimension estimates of
an optimal number of green valley galaxies required for the
analysis. It may be argued that our volume limited sample consists of
more massive galaxies which may partly explain the absence of the
environmental trends in the green valley population. However, we note
that the fraction of the red galaxies and their properties are
sensitive to their environments despite having a similar mass
range. This indicates that the quenching in the green valley galaxies
may not be driven by their environments unlike their counterparts in
the red sequence.

Our results are consistent with some recent studies with the green
valley galaxies. Mendez et al. \citep{mendez} analyzed the AEGIS data
over the redshift range $0.4<z<1.2$ and find that majority of the
galaxies in the green valley are massive disk galaxies with high
concentrations. The green galaxies are not dominated by
ongoing-mergers or post-mergers. They also pointed out that AGN-driven
winds are insufficient to explain the suppression of star formation in
the green galaxies. Recently Belfiore et al. \citep{belfiore} study
the green valley galaxies using the SDSS IV MaNGA and find that slow
quenching mechanisms which can affect the entire galaxy are needed to
explain the transition in this quasistatic population. Coertese
 \citep{cortese}analyze the GALEX and WISE data to find that at high
masses, UV colours are superior to optical colours in distinguishing
actively star-forming galaxies from passive galaxies. Another recent
paper by Angthopo et al. \citep{angthopo1} define green galaxies using
$4000\AA$ break strengh and a subsequent analysis \citep{angthopo2}
with this definition suggests that colour based selection of green
valley galaxies produces a complex population mixture caused by dust
atteneuation and the related correction. We plan to address these
issues in future works.

We finally conclude that environmental influences play a minor role
and internal processes play the dominant role in quenching star
formation in the green valley galaxies.

\section*{ACKNOWLEDGEMENT}
We thank the anonymous reviewer for the valuable comments and
suggestions that helped us to improve the manuscript. The authors
thank the SDSS and Galaxy Zoo team for making the data publicly
available. We also greatly acknowledge the efforts of the Galaxy Zoo
and Galaxy Zoo 2 volunteers for the detailed visual morphological
classifications of the SDSS galaxies.

BP would like to acknowledge financial support from the SERB, DST,
Government of India through the project CRG/2019/001110. BP would also
like to acknowledge IUCAA, Pune for providing support through
associateship programme.

Funding for the SDSS and SDSS-II has been provided by the Alfred
P. Sloan Foundation, the Participating Institutions, the National
Science Foundation, the U.S. Department of Energy, the National
Aeronautics and Space Administration, the Japanese Monbukagakusho, the
Max Planck Society, and the Higher Education Funding Council for
England. The SDSS website is http://www.sdss.org/.

The SDSS is managed by the Astrophysical Research Consortium for the
Participating Institutions. The Participating Institutions are the
American Museum of Natural History, Astrophysical Institute Potsdam,
University of Basel, University of Cambridge, Case Western Reserve
University, University of Chicago, Drexel University, Fermilab, the
Institute for Advanced Study, the Japan Participation Group, Johns
Hopkins University, the Joint Institute for Nuclear Astrophysics, the
Kavli Institute for Particle Astrophysics and Cosmology, the Korean
Scientist Group, the Chinese Academy of Sciences (LAMOST), Los Alamos
National Laboratory, the Max-Planck-Institute for Astronomy (MPIA),
the Max-Planck-Institute for Astrophysics (MPA), New Mexico State
University, Ohio State University, University of Pittsburgh,
University of Portsmouth, Princeton University, the United States
Naval Observatory, and the University of Washington.


\begin{thebibliography}{99}

\bibitem{colless01} M. Colless, et al.(for 2dFGRS team),\mnras, \textbf{328}, 1039 (2001)
\bibitem{strauss02} M.~A. Strauss, D.~H. Weinberg, R.~H. Lupton, V.~K. Narayanan, J. Annis, M. Bernardi, M. Blanton, et al., AJ, \textbf{124}, 1810 (2002)
\bibitem{strateva01} I. Strateva, {\v{Z}}. Ivezi{\'c}, G.~R. Knapp,  V.~K. Narayanan, M.~A. Strauss, J.~E. Gunn, R.~H. Lupton, et al., AJ, \textbf{122}, 1861 (2001)
\bibitem{hogg03} D.~W. Hogg, M.~R. Blanton, D.~J. Eisenstein, J.~E. Gunn, D.~J. Schlegel, I. Zehavi,  N.~A. Bahcall, et al., ApJL, \textbf{585}, L5 (2003)
\bibitem{balogh04} M.~L. Balogh, I.~K. Baldry, R. Nichol, C. Miller,  R. Bower, \&  K. Glazebrook, ApJL, \textbf{615}, L101 (2004)
\bibitem{baldry04} I.~K. Baldry, K. Glazebrook, J. Brinkmann,  {\v{Z}}. Ivezi{\'c}, R.~H. Lupton, R.~C. Nichol,\& A.~S. Szalay, ApJ, \textbf{600}, 681 (2004)
\bibitem{kauffmann03a} G. Kauffmann, T.~M. Heckman, S.~D.~M. White,  S. Charlot, C. Tremonti, E.~W. Peng, M. Seibert, et al., MNRAS, \textbf{341}, 54 (2003)
\bibitem{kannappan04}  S.~J. Kannappan, ApJL, \textbf{611}, L89 (2004)
\bibitem{wyder07} T.~K. Wyder, D.~C. Martin, D. Schiminovich, M. Seibert, T. Budav{\'a}ri , M.~A. Treyer, T.~A. Barlow, et al., ApJS, \textbf{173}, 293 (2007)
\bibitem{blanton03} M.~R. Blanton, J. Brinkmann, I. Csabai, M. Doi,  D. Eisenstein, M. Fukugita, J.~E. Gunn, et al., AJ, \textbf{125}, 2348 (2003)
\bibitem{lintott08} C.~J. Lintott, K. Schawinski, A. Slosar, et al., \mnras, \textbf{389}, 1179 (2008)
\bibitem{schawinski09a} K. Schawinski, C. Lintott, D. Thomas, M. Sarzi, D. Andreescu, S.~P. Bamford, S. Kaviraj, et al., MNRAS, \textbf{396}, 818 (2009)
\bibitem{masters10} K.~L. Masters, M. Mosleh, A.~K. Romer, R.~C. Nichol, S.~P. Bamford, K. Schawinski, C.~J. Lintott, et al., MNRAS, \textbf{405}, 783 (2010)
\bibitem{menci} N. Menci, A. Fontana, E. Giallongo, \& S. Salimbeni, ApJ, \textbf{632}, 49 (2005)
\bibitem{driver} S.~P. Driver, et al., MNRAS, \textbf{368}, 414 (2006)
\bibitem{cameron} E. Cameron, S.~P. Driver, A.~W. Graham, J. Liske, ApJ, \textbf{699}, 105 (2009)
\bibitem{cattaneo1} A. Cattaneo, A. Dekel, J. Devriendt, B. Guiderdoni, \& J. Blaizot, MNRAS, \textbf{370}, 1651 (2006)
\bibitem{cattaneo2} A. Cattaneo, et al., MNRAS, \textbf{377}, 63 (2007)
\bibitem{trayford16} J. W. Trayford, et al., MNRAS, \textbf{460}, 3925 (2016)
\bibitem{nelson} D. Nelson, et al., MNRAS, \textbf{475}, 624 (2018)
\bibitem{correa19} C.~A. Correa, J. Schaye, \& J.~W. Trayford, MNRAS, \textbf{484}, 4401 (2019)
\bibitem{bell04b} E.~F. Bell, C. Wolf, K. Meisenheimer, H.-W. Rix, A. Borch, S. Dye, M. Kleinheinrich, et al., ApJ, \textbf{608}, 752 (2004)
\bibitem{weiner05} B.~J. Weiner, A.~C. Phillips, S.~M. Faber, C.~N.~A. Willmer, N.~P. Vogt, L. Simard, K. Gebhardt, et al., ApJ, \textbf{620}, 595 (2005)
\bibitem{kriek08} M. Kriek, A. van der Wel , P.~G. van Dokkum, M. Franx, G.~D. Illingworth, ApJ, \textbf{682}, 896 (2008)
\bibitem{brammer09} G.~B. Brammer, K.~E. Whitaker, P.~G. van Dokkum, D. Marchesini, I. Labb{\'e} , M. Franx, M. Kriek, et al., ApJL, \textbf{706}, L173 (2009)
\bibitem{faber07} S.~M. Faber, C.~N.~A. Willmer, C. Wolf, D.~C. Koo, B.~J. Weiner, J.~A. Newman, M. Im, et al., ApJ, \textbf{665}, 265 (2007)
\bibitem{madau96} P. Madau, H.~C. Ferguson, M.~E. Dickinson, M. Giavalisco, C.~C. Steidel, \& A. Fruchter, MNRAS, \textbf{283}, 1388 (1996)
\bibitem{reesostriker77} M.~J. Rees, \& J.~P. Ostriker, MNRAS, \textbf{179}, 541 (1977)
\bibitem{silk77} J. Silk, ApJ, \textbf{211}, 638 (1977)
\bibitem{white78} S.~D.~M. White, \& M.~J. Rees, \mnras, \textbf{183}, 341 (1978)
\bibitem{fall80} S.~M. Fall, \& G. Efstathiou, MNRAS, \textbf{193}, 189 (1980)
\bibitem{binney04} J. Binney, MNRAS, \textbf{347}, 1093 (2004)
\bibitem{birnboim03} Y. Birnboim, \& A. Dekel, MNRAS, \textbf{345}, 349 (2003)
\bibitem{dekel06} A. Dekel, \&  Y. Birnboim, MNRAS, \textbf{368}, 2 (2006)
\bibitem{keres05} D. Kere{\v{s}} , N. Katz, D.~H. Weinberg, \& R. Dav{\'e} , MNRAS, \textbf{363}, 2 (2005)
\bibitem{gabor10} J.~M. Gabor, R. Dav{\'e} , K. Finlator, \& B.~D. Oppenheimer, MNRAS, \textbf{407}, 749 (2010)
\bibitem{gabor15} J.~M. Gabor, \& R. Dav{\'e}, MNRAS, \textbf{447}, 374 (2015)
\bibitem{birnboim07} Y. Birnboim, A. Dekel, \& E. Neistein, MNRAS, \textbf{380}, 339 (2007)
\bibitem{croton06} D.~J. Croton, V. Springel, S.~D.~M. White, G. De Lucia , C.~S. Frenk, L. Gao, A. Jenkins, et al., MNRAS, \textbf{365}, 11 (2006)
\bibitem{bower06} R.~G. Bower, A.~J. Benson, R. Malbon, J.~C. Helly, C.~S. Frenk, C.~M. Baugh, S. Cole, et al., MNRAS, \textbf{370}, 645 (2006)
\bibitem{somerville08} R.~S. Somerville, P.~F. Hopkins, T.~J. Cox, B.~E. Robertson, \& L. Hernquist, MNRAS, \textbf{391}, 481 (2008)
\bibitem{dekel08} A. Dekel, \& Y. Birnboim, MNRAS, \textbf{383}, 119 (2008)
\bibitem{dekel09b} A. Dekel, R. Sari, \& D. Ceverino, ApJ, \textbf{703},  785 (2009)
\bibitem{haywood16} M. Haywood, M.~D. Lehnert, P. Di Matteo , O. Snaith, M. Schultheis, D. Katz, \& A. G{\'o}mez, A\&A, \textbf{589}, A66 (2016)
\bibitem{spinoso17} D. Spinoso, S. Bonoli, M. Dotti, L. Mayer, P. Madau, \& J. Bellovary, MNRAS, \textbf{465}, 3729 (2017)
\bibitem{james18} P.~A. James, \& S.~M. Percival, MNRAS, \textbf{474}, 3101 (2018)
\bibitem{george19} K. George, S. Subramanian, \& K.~T. Paul, A\&A, \textbf{628}, A24 (2019)
\bibitem{combes81} F. Combes, \& R.~H. Sanders, A\&A, \textbf{96}, 164 (1981)
\bibitem{debattista04} V.~P. Debattista, C.~M. Carollo, L. Mayer, \& B. Moore, ApJL, \textbf{604}, L93 (2014)
\bibitem{kormendy04} J. Kormendy, \& R.~C. Kennicutt, ARA\&A, \textbf{42}, 603 (2004)
\bibitem{athanassoula13} E. Athanassoula, R.~E.~G. Machado, \& S.~A. Rodionov, MNRAS, \textbf{429}, 1949 (2013)
\bibitem{martig09} M. Martig, F. Bournaud, R. Teyssier, \& A. Dekel, ApJ, \textbf{707}, 250 (2009)
\bibitem{bruce16} V.~A. Bruce, J.~S. Dunlop, A. Mortlock, D.~D. Kocevski, E.~J. McGrath, \& D.~J. Rosario, MNRAS, \textbf{458}, 2391 (2016)
\bibitem{combesgerin85} F. Combes, \& M. Gerin, A\&A, \textbf{150}, 327 (1985)
\bibitem{fang13}  J.~J. Fang, S.~M. Faber, D.~C. Koo, \& A. Dekel, ApJ, \textbf{776}, 63 (2013)
\bibitem{toomre72} A. Toomre, \& J. Toomre, ApJ, \textbf{178}, 623 (1972)
\bibitem{barnes02} J.~E. Barnes, MNRAS, \textbf{333}, 481 (2002)
\bibitem{cox04} T.~J. Cox, J. Primack, P. Jonsson, \& R.~S. Somerville, ApJL, \textbf{607}, L87 (2004)
\bibitem{murray05} N. Murray, E. Quataert, \& T.~A. Thompson, ApJ, \textbf{618}, 569 (2005)
\bibitem{springel05} V. Springel, T. Di Matteo, \& L. Hernquist, MNRAS, \textbf{361}, 776 (2005)
\bibitem{gunn72} J.~E. Gunn, \& J.~R. Gott, ApJ, \textbf{176}, 1 (1972)
\bibitem{moore96} B. Moore, N. Katz, G. Lake, A. Dressler, \& A. Oemler, Nature, \textbf{379}, 613 (1996)
\bibitem{moore98} B. Moore, G. Lake, \& N. Katz, ApJ, \textbf{495}, 139 (1998)
\bibitem{balogh00} M.~L. Balogh, J.~F. Navarro, \& S.~L. Morris, ApJ, \textbf{540}, 113 (2000)
\bibitem{larson80} R.~B. Larson, B.~M. Tinsley, \& C.~N. Caldwell, ApJ, \textbf{237}, 692 (1980)
\bibitem{somerville99} R.~S. Somerville, \& J.~R. Primack, MNRAS, \textbf{310}, 1087 (1999)
\bibitem{kawata08} D. Kawata, \& J.~S. Mulchaey, ApJL, \textbf{672}, L103 (2008)
\bibitem{hoyle02}  F. Hoyle, et al., \apj, \textbf{580}, 663 (2002)
\bibitem{park05}  C. Park, et al., \apj, \textbf{633}, 11 (2005)
\bibitem{nandra07} K. Nandra, A. Georgakakis, C.~N.~A. Willmer, M.~C. Cooper, D.~J. Croton, M. Davis, S.~M. Faber, et al., ApJL, \textbf{660}, L11 (2007)
\bibitem{hasinger08} G. Hasinger, A\&A, \textbf{490}, 905 (2008)
\bibitem{silverman08} J.~D. Silverman, V. Mainieri, B.~D. Lehmer, D.~M. Alexander, F.~E. Bauer, J. Bergeron, W.~N. Brandt, et al., ApJ, \textbf{675}, 1025 (2008)
\bibitem{cimatti13} A. Cimatti, M. Brusa, M. Talia, M. Mignoli, G. Rodighiero, J. Kurk, P. Cassata, et al., ApJL, \textbf{779}, L13 (2013)
\bibitem{schawinski14} K. Schawinski, C.~M. Urry, B.~D. Simmons, L. Fortson, S. Kaviraj, W.~C. Keel, C.~J. Lintott, et al., MNRAS, \textbf{440}, 889 (2014)
\bibitem{lin17} Lin Y.-T., Hsieh B.-C., Lin S.-C., Oguri M., Chen K.-F., Tanaka M., Chiu I.-N., et al., ApJ, \textbf{851}, 139 (2017)
\bibitem{salim} S. Salim , SerAJ, \textbf{189}, 1. doi:10.2298/SAJ1489001S (2014)
\bibitem{coenda18} V. Coenda, H.~J. Mart{\'\i}nez, \& H. Muriel, MNRAS, \textbf{473}, 5617 (2018)
\bibitem{jian20} H.-Y. Jian, L. Lin, Y. Koyama, I. Tanaka, K. Umetsu, B.-C. Hsieh, Y. Higuchi, et al., ApJ, \textbf{894}, 125 (2020)
\bibitem{mihos94} J.~C. Mihos, \& L. Hernquist, ApJL, \textbf{431}, L9 (1994)
\bibitem{hubble36}  E.P. Hubble, The Realm of the Nebulae (Oxford University Press: Oxford), 79 (1936)
\bibitem{dressler80}  A. Dressler, \apj, \textbf{236}, 351 (1980)
\bibitem{postman84} M. Postman, \& M.~J. Geller, ApJ, \textbf{281}, 95 (1984)
\bibitem{lewis02} I. Lewis, M. Balogh, R. De Propris, W. Couch, R. Bower, A. Offer, J. Bland-Hawthorn, et al., MNRAS, \textbf{334}, 673 (2002)
\bibitem{gomez03} P.~L. G{\'o}mez , R.~C. Nichol, C.~J. Miller, M.~L. Balogh, T. Goto, A.~I. Zabludoff, A.~K. Romer, et al., ApJ, \textbf{584}, 210 (2003)
\bibitem{kauffmann04}  G. Kauffmann, S.~D.~M. White, T.~M. Heckman, et al., \mnras, \textbf{353}, 713 (2004)
\bibitem{bond96} J.~R. Bond, L. Kofman, \& D. Pogosyan, \nat, \textbf{380}, 603 (1996)
\bibitem{hahn07}  O. Hahn, C. Porciani, C.~M. Carollo, \& A. Dekel, \mnras, \textbf{375}, 489 (2007)
\bibitem{croton07} D.~J. Croton, L. Gao, \& S.~D.~M. White, MNRAS, \textbf{374}, 1303 (2007)
\bibitem{gao07} L. Gao, \& S.~D.~M. White, MNRAS, \textbf{377}, L5 (2007)
\bibitem{musso18} M. Musso, C. Cadiou, C. Pichon, S. Codis, K. Kraljic, \& Y. Dubois, MNRAS, \textbf{476}, 4877 (2018)
\bibitem{vakili19} M. Vakili, \& C. Hahn, ApJ, \textbf{872}, 115 (2019)
\bibitem{miyatake16} H. Miyatake, S. More, M. Takada, D.~N. Spergel, R. Mandelbaum, E.~S. Rykoff, \& E. Rozo, PhRvL, \textbf{116}, 041301 (2016)
\bibitem{montero17} A.~D. Montero-Dorta, E. P{\'e}rez, F. Prada, S. Rodr{\'\i}guez-Torres , G. Favole, A. Klypin, R. Cid Fernandes, et al., ApJL, \textbf{848}, L2 (2017)
\bibitem{kerscher18} M. Kerscher, A\&A, \textbf{615}, A109 (2018)
\bibitem{pandey06} B. Pandey, \&  S. Bharadwaj, \mnras, \textbf{372}, 827 (2006)
\bibitem{pandey08}  B. Pandey, \& S. Bharadwaj, \mnras, \textbf{387}, 767 (2008)
\bibitem{scudder12}  J.~M. Scudder, S.~L. Ellison, \& Mendel, J.~T., \mnras, \textbf{423}, 2690 (2012)
\bibitem{darvish14}  B. Ellison, D. Sobral, B. Mobasher, et al., \apj, \textbf{796}, 51 (2014)
\bibitem{filho15} M.~E. Filho, J. S{\'a}nchez Almeida, C.Mu{\~n}oz-Tu{\~n}{\'o}n, et al., \apj, \textbf{802}, 82 (2015)
\bibitem{luparello15} H.~E. Luparello, M. Lares, D. Paz, et al., \mnras, \textbf{448}, 1483 (2015)
\bibitem{pandey17} B. Pandey, \& S.Sarkar ,MNRAS, \textbf{467}, L6 (2017)
\bibitem{sarkar20} S. Sarkar, \& B.Pandey , MNRAS, \textbf{497}, 4077 (2020)
\bibitem{bhattacharjee20} S. Bhattacharjee, B. Pandey, \& S. Sarkar, JCAP, \textbf{2020}, 039 (2020)
\bibitem{pandey20a} B. Pandey, \& S. Sarkar, MNRAS, \textbf{498}, 6069 (2020)
\bibitem{fisher08} D.~B. Fisher, \& N. Drory, AJ, \textbf{136}, 773 (2008)
\bibitem{lang14} P. Lang, S. Wuyts, R.~S. Somerville, N.~M. F{\"o}rster Schreiber , R. Genzel, E.~F. Bell, G. Brammer et al., ApJ, \textbf{788}, 11 (2014)
\bibitem{forster14} N.~M. F{\"o}rster Schreiber, R. Genzel, S.~F. Newman, J.~D. Kurk, D. Lutz, L.~J. Tacconi, S. Wuyts, et al., ApJ, \textbf{787}, 38 (2014)
\bibitem{leslie16} S.~K. Leslie, L.~J. Kewley, D.~B. Sanders, \& N. Lee, MNRAS, \textbf{455}, L82 (2016)
\bibitem{fabian12} A.~C. Fabian, ARA\&A, \textbf{50}, 455 (2012)
\bibitem{gunn98} J.~E. Gunn, M. Carr, C. Rockosi, M. Sekiguchi, K. Berry, B. Elms,  E. de Haas, et al., AJ, \textbf{116}, 3040 (1998)
\bibitem{gunn06} J.~E. Gunn, W.~A. Siegmund, E.~J. Mannery, R.~E. Owen, C.~L. Hull, R.~F. Leger, L.~N. Carey, et al., AJ, \textbf{131}, 2332 (2006)
\bibitem{york00}  D.~G. York, et al., \aj, \textbf{120}, 1579 (2000)
\bibitem{ahumada20} R. Ahumada, C. Allende Prieto, A. Almeida, F. Anders, S.~F. Anderson, B.~H. Andrews, B. Anguiano, et al., ApJS, \textbf{249}, 3 (2020)
\bibitem{conroy09}  C. Conroy, J.~E. Gunn, \& M. White, ApJ, \textbf{699}, 486 (2009)
\bibitem{bruzal83} A.~G. Bruzual, ApJ, \textbf{273}, 105 (1983)
\bibitem{brinchmann04} J. Brinchmann, S. Charlot, S.~D.~M. White, C. Tremonti, G. Kauffmann, T. Heckman, \& J. Brinkmann, MNRAS, \textbf{351}, 1151 (2004)
\bibitem{kauffmann03b} G. Kauffmann, T.~M. Heckman, S.~D.~M. White, S. Charlot, C. Tremonti, J. Brinchmann, G. Bruzual, et al., MNRAS, \textbf{341}, 33 (2003)
\bibitem{gandalf} M. Sarzi, J. Falc{\'o}n-Barroso, R.~L.Davies, R. Bacon, M. Bureau, M. Cappellari, P.~T. de Zeeuw  et al., MNRAS, \textbf{366}, 1151 (2006)
\bibitem{ppxf} M. Cappellari, E. Emsellem, PASP, \textbf{116}, 138 (2004)
\bibitem{planck18} Planck Collaboration, N. Aghanim, Y. Akrami, M. Ashdown, J. Aumont, C. Baccigalupi, M. Ballardini, et al., A\&A, \textbf{641}, A6 (2018)
\bibitem{pandey20b}  B. Pandey,MNRAS, \textbf{499}, L31 (2020)
\bibitem{zadeh} L. A. Zadeh,Fuzzy sets. Information and Control, \textbf{8}, 338 (1965)
\bibitem{willett} K.~W. Willett, C.~J. Lintott, S.~P. Bamford, K.~L. Masters, B.~D. Simmons, K.~R.~V. Casteels, E.~M. Edmondson, et al., MNRAS, \textbf{435}, 2835 (2013)
\bibitem{sarkar09} P. Sarkar, \& S. Bharadwaj, \mnras, \textbf{394}, L66 (2009)
\bibitem{sarkar19}  S. Sarkar, \& B. Pandey, MNRAS, \textbf{485}, 4743 (2019)
\bibitem{alpaslan14} M. Alpaslan, et al., MNRAS, \textbf{438}, 177 (2014)
\bibitem{peng} Y.-.jie Peng,  S.~J. Lilly, K. Kova{\v{c}}, M. Bolzonella,  L. Pozzetti,  A. Renzini, G. Zamorani, et al., ApJ, \textbf{721}, 193 (2010)
\bibitem{kaviraj09} S. Kaviraj, S. Peirani, S. Khochfar, J. Silk, \& S. Kay, MNRAS, \textbf{394}, 1713 (2009)
\bibitem{thomas10} D. Thomas, C. Maraston, K. Schawinski, M. Sarzi, \& J. Silk, MNRAS, \textbf{404}, 1775 (2010)
\bibitem{schawinski09b}  K. Schawinski, S. Virani,  B. Simmons, C.~M. Urry,  E. Treister, S. Kaviraj, B. Kushkuley, ApJL, \textbf{692}, L19 (2009)
\bibitem{schawinski15}  K. Schawinski, M. Koss, S. Berney, L.~F. Sartori, MNRAS, \textbf{451}, 2517 (2015)
\bibitem{tacchella} S. Tacchella, B. Diemer, L. Hernquist,  S. Genel, F. Marinacci, D. Nelson,  A. Pillepich, et al., MNRAS, \textbf{487}, 5416 (2019)
\bibitem{mendez} A.~J. Mendez, A.~L. Coil, J. Lotz, S. Salim, J. Moustakas, \& L. Simard, ApJ, \textbf{736}, 110 (2011)
\bibitem{belfiore} F. Belfiore, R. Maiolino,  K. Bundy, K. Masters, M. Bershady, G.~A. Oyarz{\'u}n, L. Lin , et al., MNRAS, \textbf{477}, 3014 (2018)
\bibitem{cortese} L. Cortese, A\&A, \textbf{543}, A132 (2012)
\bibitem{angthopo1} J. Angthopo, I. Ferreras, J. Silk, MNRAS, \textbf{488}, L99 (2019)
\bibitem{angthopo2} J. Angthopo, I. Ferreras, J. Silk, MNRAS, \textbf{495}, 2720 (2020)









\end{thebibliography}
\end{document}